\documentclass[journal]{IEEEtran}

\usepackage{color,array,amsthm}
\usepackage{graphicx}

\usepackage{titlesec}
\titleformat{\subsubsection}[runin]  
  {\itshape}                       
  {\hspace{1em}\arabic{subsubsection})}  
  {0.5em}                            
  {}                               
  [\hspace{0em}:]                

\titlespacing*{\subsubsection}{0em}{*0}{0.5em}  

\usepackage[linesnumbered,ruled,vlined]{algorithm2e}
\SetKwInput{KwInput}{Input}                
\SetKwInput{KwOutput}{Output}              
\usepackage{amsfonts}
\usepackage{cite}
\usepackage{amsmath,amssymb,amsfonts}
\usepackage{algorithmicx}
\usepackage{textcomp}
\usepackage{multirow}
\usepackage{graphicx}
\usepackage[table,xcdraw]{xcolor}
\usepackage{comment}
\usepackage{minted}
\usepackage{amsthm}
\usepackage{url} 
\usepackage{algcompatible}

\newcommand{\USCKC}{\ensuremath{\sf USCKC}}
\newcommand{\PH}{\text{PH}}
\newcommand{\AC}{\text{AC}}
\newcommand{\TA}{\text{TA}}
\newcommand{\TE}{\text{TE}}
\renewcommand{\S}{\text{\bf S}}
\newcommand{\G}{\text{\bf G}}
\newcommand{\U}{\text{\bf U}}
\renewcommand{\L}{\text{\bf L}}
\newcommand{\BS}{\text{\sf BS}}
\newcommand{\PL}{\text{\sf PL}}
\newcommand{\GS}{\text{\sf GS}}
\newcommand{\MC}{\text{\sf MC}}
\newcommand{\DPC}{\text{\sf DPC}}
\newcommand{\RT}{\text{\sf RT}}
\renewcommand{\SS}{\text{\bf SS}}
\newcommand{\GG}{\text{\bf GG}}
\newcommand{\SG}{\text{\bf SG}}
\newcommand{\SU}{\text{\bf SU}}
\newcommand{\GU}{\text{\bf GU}}
\newcommand{\UU}{\text{\bf UU}}
\newcommand{\C}{\text{\sc C}}
\newcommand{\I}{\text{\sc I}}
\newcommand{\A}{\text{\sc A}}

\newcommand{\AvailableCapabilitySet}{\text{\sc JAT}}

\newcommand{\ActivitySet}{\text{\sc ACT}}
\newcommand{\AttackDescription}{\text{\sc AD}}

\newcommand{\In}{\ensuremath{\sf in}}
\newcommand{\Through}{\ensuremath{\sf through}}
\newcommand{\Out}{\ensuremath{\sf out}}

\newcommand{\ignore}[1]{}

\usepackage{xcolor} 
\definecolor{LightGray}{gray}{0.9}
\newtheorem{insight}{Insight}
\newtheorem{definition}{Definition}
\definecolor{LightGray}{gray}{0.97}
\def\BibTeX{{\rm B\kern-.05em{\sc i\kern-.025em b}\kern-.08em
    T\kern-.1667em\lower.7ex\hbox{E}\kern-.125emX}}

\IEEEoverridecommandlockouts

\begin{document}

\title{Characterizing Cyber Attacks against Space Infrastructures with Missing Data: Framework and Case Study\thanks{The conference version of this paper appeared as \cite{ear2023CNS}}
}

\markboth{EAR ET AL. CHARACTERIZING CYBER ATTACKS AGAINST SPACE INFRASTRUCTURES}{}

\author{\IEEEauthorblockN{Ekzhin Ear, Jose Luis Castanon Remy,
Caleb Chang,
Qiren Que, Antonia Feffer
and 
Shouhuai Xu
}
\IEEEauthorblockA{\{eear, jcastano, cchang, qque2, afeffer2, sxu\}@uccs.edu\\
Laboratory for Cybersecurity Dynamics, Department of Computer Science, University of Colorado Colorado Springs}}

\maketitle

\begin{abstract}Cybersecurity of space infrastructures is an emerging topic,
despite space-related cybersecurity incidents occurring as early as 1977 (i.e., hijacking of a satellite transmission signal). 
There is no single dataset that documents cyber attacks against space infrastructures that have occurred in the past; instead, these incidents are often scattered in media reports while missing many details, which we dub the {\em missing-data} problem. Nevertheless, even ``low-quality” datasets containing such reports would be extremely valuable because of the dearth of space cybersecurity data and the sensitivity of space infrastructures which are often restricted from disclosure by governments. This prompts a research question: How can we characterize real-world cyber attacks against space infrastructures? In this paper, we address the problem by proposing a framework, including metrics, while also addressing the missing-data problem 
by leveraging methodologies such as the Space Attack Research and Tactic Analysis (SPARTA) and the Adversarial Tactics, Techniques, and Common Knowledge (ATT\&CK) to 
``extrapolate” the missing data in a principled fashion.
We show how the extrapolated data can be used to reconstruct ``hypothetical but plausible'' {\em space cyber kill chains} and {\em space cyber attack campaigns} that have occurred in practice. 
To show the usefulness of the framework, we extract data for 108 cyber attacks against space infrastructures and show how to extrapolate this ``low-quality” dataset containing missing information to derive 6,206
attack technique-level space cyber kill chains. Our findings include: cyber attacks against space infrastructures are getting increasingly sophisticated; 
successful protection of the link segment between the space and user segments could have thwarted nearly half of the 108 attacks.
We will make our dataset available.

\end{abstract}



\section{INTRODUCTION}\label{sec:intro}

Space infrastructures have become an underpinning of modern society because they provide services that support many land, air, maritime, and cyber operations, such as Positioning, Navigation, and Timing (PNT) for the global stock market \cite{madry2015applications}, Satellite Communications (SATCOM) for global beyond-line-of-sight (BLOS) terrestrial voice and data requirements, and remote sensing for space domain awareness and planetary defense (e.g., detecting and deflecting large debris from hitting Earth). At a high level, space infrastructures 
can be understood via their 4 segments: space (e.g., satellites), ground (e.g., radar facilities), user (e.g., radio receivers), and link (e.g., radio frequency signals between the other segments). 
Moreover, space infrastructures can be described at multiple levels of abstraction, such as the mission, segment, and component levels described in  \cite{XuSpaceSec2025,XuS&P2025}. 

The space domain is interwoven with and enabled by the cyber domain, with real-time
cyber-physical systems comprising the space segment. Consequently, cyber attacks can affect
space infrastructures, as evidenced by numerous space-related security incidents \cite{SpaceSecurityInfo, pavur2022building, fritz2013satellite, falco2021security}. 
Space incidents have occurred as early as 1977, with the hijacking of a satellite's audio transmission to broadcast the
attacker's own message \cite{fritz2013satellite}. In 1998, a U.S.-German RoSat (sensing satellite) experienced a malfunction that led to the satellite turning its x-ray sensor 
towards the sun, causing
permanent damage.
While it is debatable whether this incident was caused by cyber attacks,
the confirmed cyber attack
at the Goddard Space Flight Centre, where the RoSat is controlled, 
shows that cyber attacks can cause physical damage to space infrastructures \cite{Wess2021}. 
Although cyber attacks against space infrastructures have become a reality, there is no systematic understanding of  real-world cyber attacks against space infrastructures, likely due to the lack of data.

In this paper, we initiate the study of the problem in characterizing {\em real-world} cyber attacks against space infrastructures {\em with missing data}.
The problem is important to deepen our understanding of cyber attacks against space infrastructures, which are not yet understood, 
and to gain insights into making future
space infrastructures secure. 
It would be ideal that we have well documented attacks with significantly detailed descriptions (e.g., through digital forensics) to 
serve as input to this characterization study.
Unfortunately, cyber attacks, especially those against space infrastructures, are rarely well-documented owing to a variety of reasons (e.g., sensitivity), which explains why we must embrace missing data. 

Note that the perspective of our study is different from textbook cyber threat models, which typically {\em assume} what an attacker attempts to achieve and what capabilities are available to an attacker. Moreover, a competent design is often able to thwart the attacks specified in well-defined threat models; for instance, a cryptographic protocol can be rigorously proven to thwart the attacks specified in a rigorously defined threat model. However, the fact that attacks succeed in the real world means that the threat model considered in the design phase is incomplete and/or the design (of the employed defense)
is not competent. 

\smallskip

\noindent{\bf Our contributions}.
We make two technical contributions. 
First, we propose a novel framework to characterize real-world cyber attacks against space infrastructures {\em with missing data}. The framework has three characteristics: 
(i) It is {\em general} because it can accommodate both cyber attacks against space infrastructures that occurred in the past and cyber attacks that may occur in the future. 
(ii) It is {\em practical} because it explicitly deals with missing details of attacks, which is often the case with real-world datasets.
At a high level, we achieve this by leveraging the Aerospace Space Attack Research and Tactic Analysis (SPARTA) \cite{SPARTA} and the MITRE Adversarial Tactics, Techniques, and Common Knowledge (ATT\&CK) \cite{ATTCK} frameworks to extrapolate the missing details of attacks.
(iii) It offers three {\em metrics} for measuring {\em attack consequence}, {\em attack sophistication}, and {\em Unified Space Cyber Kill Chain ($\USCKC$) likelihood}, where the concept of $\USCKC$ will be elaborated later.
These metrics might be of independent value because they could be adapted to other kinds of infrastructures and networks.

Second, we show the usefulness of the framework by applying it to characterize the cyber attacks described in a dataset, which is prepared by this study and is, to our knowledge, the first comprehensive dataset of cyber attacks against space infrastructures. 
The dataset includes 108 attacks, which do not include the attacks studied in the academic literature (e.g.,
\cite{pavur2020tale,XuS&P2025}) because these attacks are not known having occured in the real world.
Among the 108 attacks, 72 are manually extracted from four publicly available datasets documenting space-related incidents, noting that these four datasets are not geared toward space cyber attacks; the 72 attacks are described in the conference version of the present paper, namely \cite{ear2023CNS}, 
and the other 36 attacks are obtained via our own Internet search and analysis.
Since the description for each of these 108 attacks, as obtained from the Internet, contains missing data, especially regarding the details of the attack, we must extrapolate the missing data in order to speculatively reconstruct the attack. This extrapolation leads to a total of 6,206 probable {\em attack technique}-level $\USCKC$s for the 108 attacks, which highlights the high uncertainty incurred by the missing details in attack description, noting that we use the standard concepts of attack tactic and attack technique described in \cite{strom2018mitre}.
This allows us to draw a number of insights, such as: attacks against space infrastructures can be effectively mitigated by hardening the ground segment; average-sophisticated cyber attacks have been effective against space infrastructures; attacks are getting increasingly sophisticated, perhaps because of the increasing employment of defenses; 
defending the links  between the space segment and the user segment could have thwarted nearly half of the 108 attacks. We will make this dataset publicly available.

\smallskip

\noindent{\bf Paper outline}.
Section \ref{sec:termsAndConcepts} describes terms and concepts used throughout the paper.
Section \ref{sec:framework} presents the framework.
Section~\ref{sec:casestudy} presents a case study.
Section~\ref{sec:limits} discusses limitations of the present study. 
Section~\ref{sec:related_works} reviews prior studies. 
Section~\ref{sec:conclusion} concludes the paper.

\section{TERMINOLOGIES AND CONCEPTS}\label{sec:termsAndConcepts}

\noindent{\bf Cyber Attack Campaign}. A cyber attack campaign is an attacker's activities against a (space) infrastructure or network in attempting to accomplish certain objectives. These  activities are typically manifested as multi-step attacks. The first attack step typically compromises a space infrastructure unit (at an appropriate level of abstraction), dubbed {\em entry node}, whereby the attacker penetrates into the space infrastructure. The last attack step typically compromises another unit (at the same level of abstraction), dubbed {\em objective node}, where the attacker accomplishes its overall objectives. Each objective is often divided into  multiple sub-objectives. 

\smallskip

\noindent{\bf ATT\&CK}. We adopt the following terms from ATT\&CK \cite{strom2018mitre}:
an {\it attack tactic} specifies {\em what} an attacker attempts to accomplish (typically a sub-objective); an {\it attack technique} specifies {\em how} the attacker attempts to accomplish an attack tactic;
an {\it attack procedure} is a concrete implementation of 
an attack technique, noting that an attack technique could have many concrete implementations or instantiations. 
At a high level, ATT\&CK describes possible attack tactics, techniques, and procedures against enterprise IT networks \cite{ATTCK,strom2018mitre}. 
We use ${\rm ATT}_{\rm TA}$ to denote the set of ATT\&CK tactics, meaning ${\rm ATT}_{\rm TA}=\{$Reconnaissance, Resource Development, Initial Access, Execution, Persistence, Privilege Escalation, Defense Evasion, Credential Access, Discovery, Lateral Movement, Collection, Command \& Control, Exfiltration, Impact$\}$. We use ${\rm ATT}_{\rm TE}$ to denote the set of ATT\&CK techniques, which are too many to be listed here but can be found at  \cite{strom2018mitre}. 
Real-world use cases of ATT\&CK include: Cyber Threat Intelligence (CTI) analysts use it to extract useful information from raw data
\cite{parmar2019use}; digital forensics and incident response analysts use it to characterize cyber incidents \cite{virkud2024does}. 

\smallskip

\noindent{\bf Cyber Kill Chain and Unified Kill Chain (UKC)}. The concept of {\em Cyber Kill Chain}
\cite{Lockheed} aims to help see the ``big picture'' of cyber attack campaigns. Attack tactics can be chained together to formulate tactic-level Cyber Kill Chains, and attack techniques can be chained together to formulate technique-level Cyber Kill Chains
\cite{Lockheed}.
Both tactic- and technique-level Cyber Kill Chains present an abstract description of 
cyber attack campaigns.
The concept of Unified Kill Chain (UKC) \cite{pols2017unified} improves upon the concept of Cyber Kill Chain by providing three major phases of a cyber campaign: access into a victim network, denoted by ``$\In$''; pivoting through the victim network, denoted by ``$\Through$''; and achieving overall objectives of the attack, denoted by
``$\Out$''. 
This is reasonable because a cyber attack campaign
typically begins with an $\In$ phase, followed by one or multiple 
$\Through$ phases, and concluding with an $\Out$ phase. However, some attack campaigns may not have all the three phases.

\smallskip

\noindent{\bf SPARTA}. Since ATT\&CK is geared towards enterprise IT networks, it is not well suited for space infrastructures and systems. This motivates the introduction of SPARTA \cite{SPARTA}, which describes attack tactics, techniques, and procedures 
against space infrastructures and systems, especially for those which have no counterpart in enterprise IT networks, such as attacks against satellites and satellite-ground communications.
We use ${\rm SPA}_{\rm TA}$ to denote the set of SPARTA tactics, meaning ${\rm SPA}_{\rm TA}=\{$Reconnaissance, Resource Development, Initial Access, Execution, Persistence, Defense Evasion, Lateral Movement, Exfiltration, Impact$\}$. Note that ${\rm SPA}_{\rm TA}\subset {\rm ATT}_{\rm TA}$. We use ${\rm SPA}_{\rm TE}$ to denote the set of SPARTA techniques, which are too many to be listed here but can be found at  \cite{SPARTA}. 
Real-world use cases of SPARTA 
are similar to those of ATT\&CK's. 
In principle, SPARTA-based tactics and techniques can be chained together as in the case of ATT\&CK. However, the resulting chains may not be sufficient because SPARTA is geared towards space infrastructures {\em without} the ground systems. This explains why we propose incorporating both ATT\&CK and SPARTA to formulate more comprehensive kill chains.






\smallskip

\noindent{\bf Unified Space Cyber Kill Chain ($\USCKC$).}
We propose adapting the concept of UKC to the concept of $\USCKC$, which cuts across all the 4 segments of space infrastructures (including the space segment and the ground segment), justifying our proposal of leveraging SPARTA and ATT\&CK. As a result, we can see the ``big picture'' of space cyber attack campaigns that may be waged from ground systems but target the space segment. Moreover, SPARTA and ATT\&CK attack tactics (or techniques) can be chained together to formulate tactic-level (or technique-level) $\USCKC$s,
which present abstract descriptions of space cyber attack campaigns. To unify notations, we denote by
$\AvailableCapabilitySet_{\TA} = {\rm SPA}_{\rm TA}\bigcup {\rm ATT}_{\rm TA}= {\rm ATT}_{\rm TA}$ the set of {\em joint SPARTA-ATT\&CK attack tactics} while recalling ${\rm SPA}_{\rm TA}\subset {\rm ATT}_{\rm TA}$, and by $\AvailableCapabilitySet_{\TE} = {\rm SPA}_{\rm TE}\bigcup {\rm ATT}_{\rm TE}$ the set of {\em joint SPARTA-ATT\&CK attack tactics}. 

\ignore{
A space cyber kill chain is composed of (i) a {\it space cyber attack tactic kill chain}; and (ii) $m$ {\it space cyber attack technique kill chains}. A {\it space cyber attack tactic kill chain} is a vector of attack tactics $\AvailableCapabilitySet_{\TA}$ the attacker may use against space infrastructures. A {\it space cyber attack technique kill chain} is a vector of attack techniques $\AvailableCapabilitySet_{\TE}$ the attacker may use against a space infrastructure modules to accomplish an {\it attack tactic} in $\AvailableCapabilitySet_{\TA}$.
}

\smallskip

\noindent{\bf Reconstructing $\USCKC$s via Activities}. At a high level, a $\USCKC$ is described by an $\In$ phase, one or multiple $\Through$ phases, and an $\Out$ phase, while noting that uncommon attacks may have multiple $\In$ and/or $\Out$ phases. Our goal is to reconstruct $\USCKC$s from CTI reports on space cyber attack campaigns. To accomplish this, we adapt the following concepts from the digital forensics community \cite{iqbal2020digital}. 
\begin{itemize}
\item {\it Objective activities}: Two 
attack tactics in $\AvailableCapabilitySet_{\TA}$, {\em Exfiltration} and {\em Impact}, are objective activities because they describe the overall impact of a space cyber attack campaign. 
\item {\it Milestone activities}: Three attack tactics in $\AvailableCapabilitySet_{\TA}$, 
{\em Initial Access}, {\em Lateral Movement}, and {\em Credential Access}, are milestone activities because they advance
an attack campaign 
towards an objective activity.
\item {\it Enabling activities}: Six attack tactics in $\AvailableCapabilitySet_{\TA}$,
{\em Resource Development}, {\em Execution}, {\em Privilege Escalation}, {\em Persistence}, {\em Command \& Control}, and {\em Defense Evasion}, are enabling activities because they seek to establish or modify a current state of a system environment to facilitate an objective or milestone activity.
\item {\it Information discovery activities}: Three attack tactics in $\AvailableCapabilitySet_{\TA}$, 
{\em Reconnaissance}, {\em Discovery}, and {\em Collection}, are information discovery activities because they seek to provide necessary information to support an objective, milestone, 
or enabling activity.
\end{itemize}

\section{FRAMEWORK}\label{sec:framework}





\subsection{Framework Requirements and Overview} 


A framework for analyzing cyber attacks against space infrastructures 
should satisfy the following requirements:
\begin{itemize}
\item {\bf{Requirement 0}:} It is {\em general} enough to accommodate both past 
and future space cyber attacks.

\item {\bf Requirement 1}: It is {\em practical} by accommodating real-world datasets with missing attack details.

\item {\bf{Requirement 2}:} It provides metrics for characterizing cyber attacks against space infrastructures.
\end{itemize}
To address Requirement 0, we propose a system model that can be used to describe past and future attacks. To address Requirement 1, we ``extrapolate'' a given description of a space cyber attack to incorporate {\em hypothetical but plausible} missing details.  
To address Requirement 2, we define {\em attack consequence}, {\em attack sophistication}, and {\em attack likelihood} metrics.

\begin{figure}[!htbp]
\centering
\includegraphics[width=.9\columnwidth]{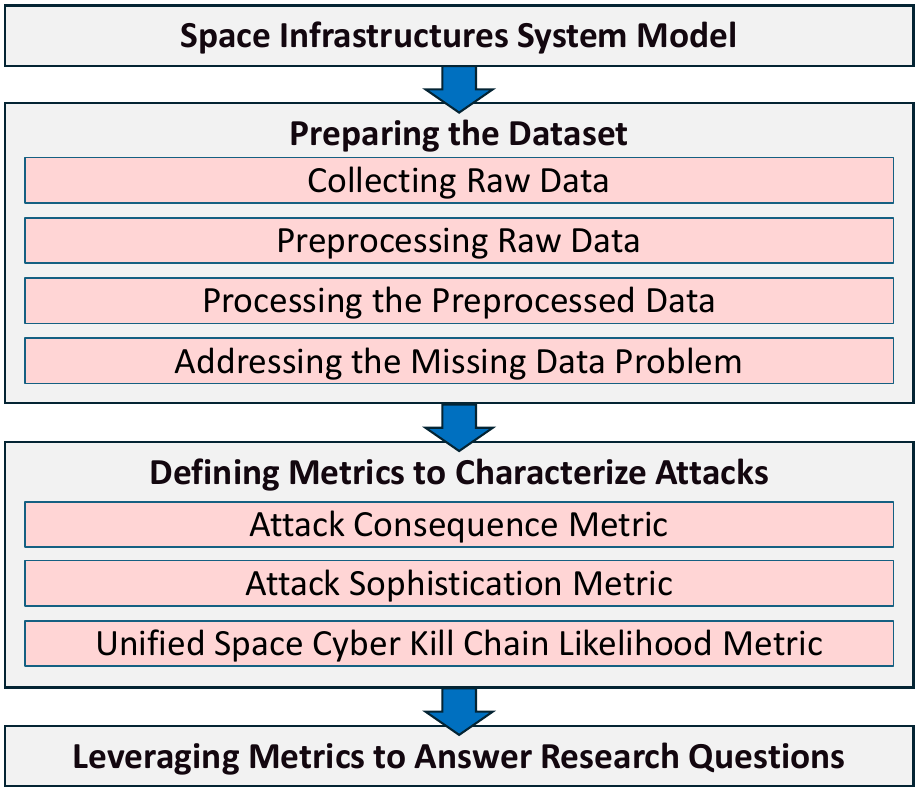}
\caption{The framework. 
}
\label{fig:framework}
\end{figure}

Figure~\ref{fig:framework} highlights the framework, which has four major components: (i) designing system model; (ii) preparing dataset; 
(iii) defining metrics to  characterize attacks; and (iv) leveraging metrics to answer research questions.

\subsection{Space Infrastructure System Model}\label{subsec:system model}

\begin{figure}[!htbp]
\centering{\includegraphics[width=\columnwidth]{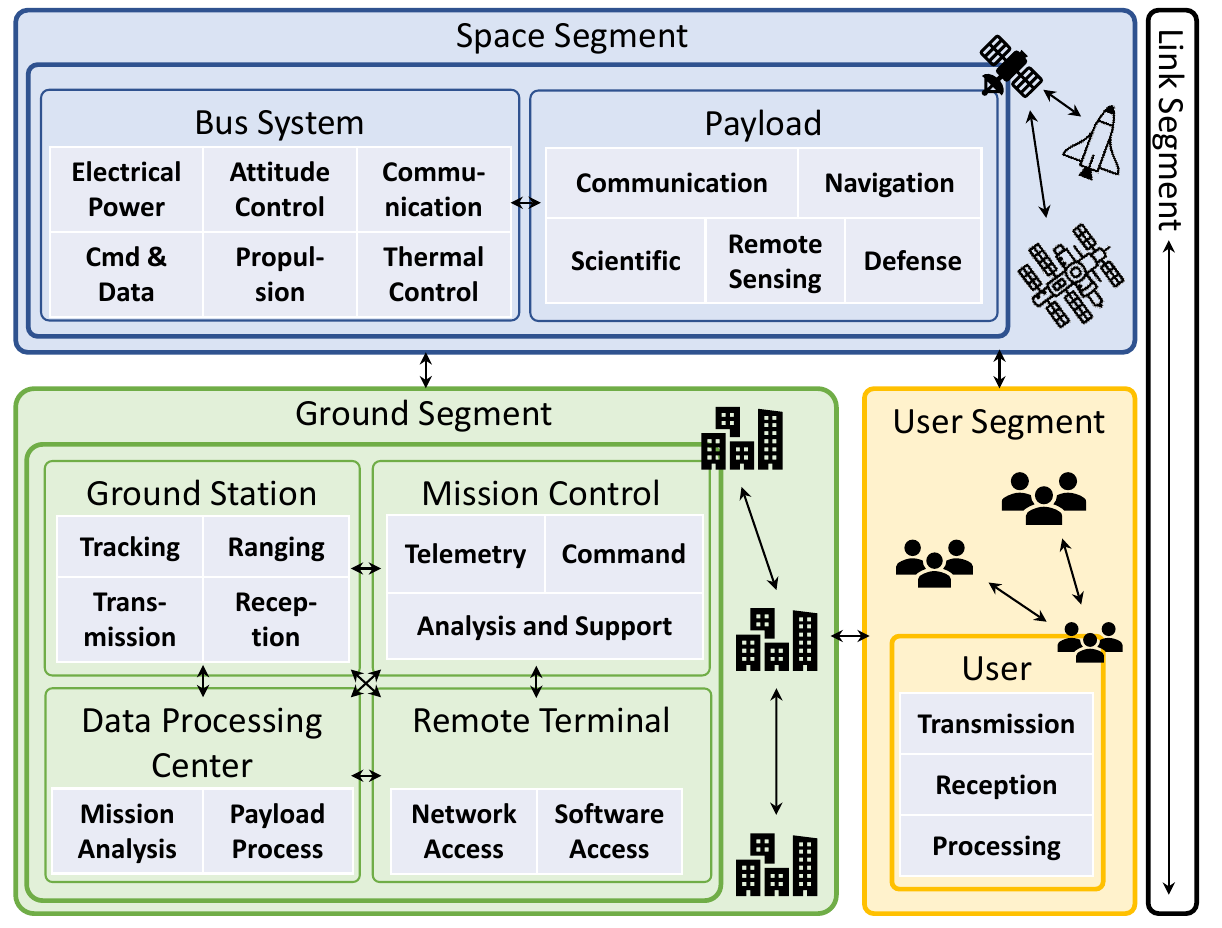}}
\vspace{-2em}
\caption{System model showing the 4 segments of space infrastructures.}
\label{fig:system_model}
\end{figure}

As mentioned above, a space infrastructure often consists of 4 segments: {\em space}, {\em link}, {\em ground}, and {\em user} \cite{tedeschi2022satellite, guo2021survey, fritz2013satellite}. This prompts us to propose the system model highlighted in Figure~\ref{fig:system_model}. 
We advocate separation between the ground segment and the user segment because the participants of each segment are different (e.g., an aerospace engineer versus a research scientist) and users often interact with a payload of space infrastructures whereas the ground segment interacts with both the payloads and the bus. Further, this separation increases the precision of cybersecurity design and analysis. For example, certain defense tools may be best employed at one segment but not the other, and cyber threat actors may desire to attack users without inflicting impact upon the ground and/or space segments.

The space segment includes satellites, spacecrafts, and space stations. 
A {\em Bus System} component facilitates  tracking, telemetry, and command requirements (TT\&C) and typically contains the following modules \cite{chippalkatti2021recent, nguyen2020future}: {\em electrical power}, {\em attitude control}, {\em communication}, {\em command and data handling}, {\em propulsion}, and {\em thermal control}.
These modules may be attacked through the cyber domain. 
The {\em Payload} component of space infrastructures include the following modules: {\em communication} (e.g., antennas and transmitters for relaying voice and data), {\em navigation} (e.g., Global Navigation Satellite System (GNSS) receivers 
for position and timing), {\em scientific experiment} (e.g., telescopes and spectrometers for research), {\em remote sensing} (e.g., sensors and cameras for terrestrial environmental monitoring), and {\em defense} (e.g., national security or military equipment for reconnaissance). 
Note that some space infrastructures may employ legacy technologies because they often cannot be replaced after being launched into the space. Note also that services such as data centers may be established in space in the future especially because of the increasingly low launch costs \cite{soesanto2021terra}.

In the ground segment, the {\em Ground Station} component contains the hardware and software for transmitting/receiving RF signals and  tracking/ranging objects in space. The {\em Mission Control} component processes telemetry data to assess the health of modules in the space segment, sends commands to control the modules 
in the space segment, conducts analyses to plan orbital maneuvers and assess conjunctions, and manages other aspects of space infrastructure operations. The {\em Data Processing Center} component conducts deeper analyses of space infrastructure missions and processes payload data. The {\em Remote Terminal} component provides a light-weight software stack and network connectivity to the other elements of the ground segment.
The ground segment often reflects the architecture of traditional IT enterprise networks, with the addition of space-specific hardware and software, such as satellite terminals, modems, and flight software. The increasing commercialization of space acquisition has caused the supply chain for space-specific systems to trend towards enterprise network requirements. The geographical dispersion of ground stations presents unique challenges because stations across the globe have varying degrees of 
physical security \cite{soesanto2021terra}.

The user segment is also geographically dispersed across continents and oceans, often requiring space infrastructure services around the clock.
GNSS is an example of such service used across the land, air, and maritime domains in automobiles, airplanes, and ships across the world. The user segment typically receives data directly from GNSS satellites while SATCOM (Satellite Communications) users also transmit traffic to satellites. 
At times, the user segment may process data, such as researchers processing images from scientific satellites hosting telescopes.

The link segment is concerned with inter-segment (e.g., satellite to ground) and intra-segment (i.e., satellite to satellite) data connections. Satellites communicate by passing data through the link segment with other satellites, ground stations, as well as directly to user terminals, potentially via its payload (e.g., for voice and data transfer)  or bus (for TT\&C). It possesses a variety of physical and electromagnetic properties across the spectrum. For our purpose, it suffices to consider the link segment from a cyber perspective, namely the data transiting the link segment, rather than the physical properties of the link segment. 
For example, satellites regularly communicate with the ground and user segments, as well as other satellites, for TT\&C and mission execution of payloads. Elements of the ground segment may share data with each other as part of the same facility or across wide area networks, as well as forward and retrieve data to the user segment. Members of the user segment may also collaborate with other users, e.g., scientists processing data or internet service providers broadcasting packets.



We represent a space infrastructure at the module level of abstraction, as a directed graph $G=(V,E)$, where $V$ is the set of nodes representing space infrastructure modules, and $E$ is the set of arcs representing communication relationships such that $(v_1,v_2)\in E$ means $v_1$ can initiate unidirectional or bidirectional communications with $v_2$.

\subsection{Preparing Dataset}
\label{subsec:preparing_dataset}

\ignore{

We identify an innovative perspective, which is the {\em cyber threat characteristics}. 
We define this through four attributes (and accompanying metrics): impact $\to$ complexity $\to$ capability strength. 
Given our system model and the available raw real-world data described in Section~\ref{sec:context}, we consider the impacts of real-world cyber threat actors against various elements and segments of space infrastructures. These impacts imply a level of complexity that threat actors must overcome to achieve their objectives. Their ability to do this implies their capability strength, from which we can derive the minimal capability requirement for threat actors to achieve their desired impacts.

}



It would be ideal that a dataset of cyber attacks against space infrastructures possesses the following properties:
\begin{itemize}
\item {\em Comprehensiveness}: This property deals with the coverage of a dataset, in terms of the aspects that are important to describe space cyber attacks. Ideally, a dataset should contain every step of every attack, including attack tactics, techniques, and procedures so that one can reconstruct the $\USCKC$ as a useful abstraction for further studies. This means that the data source should have conducted forensics at the greatest details possible. 
Clearly, this property is hard to guarantee in the real world 
because there is a lack of deployed sensors, which also hinders the effectiveness of forensics techniques, and data about space infrastructures are typically classified by governments  (i.e., not provided to the public).

\item {\em Accuracy}: The details about attack tactics, 
techniques, and procedures described in a dataset should be accurate, leaving no room for ambiguity or misinterpretation. This property is also hard to guarantee because descriptions in real-world datasets are often ambiguous
(e.g., different people may use different terms when documenting space cyber attacks and many sources often come from news reporting rather than rigorous investigations).

\item {\em Zero Missing Data}: This means every aspect of an attack that should be described in a dataset is indeed described in the dataset. While the accuracy property implies that the provided descriptions should be accurate, 
the present property is different because a dataset can be accurate on the provided description but there can be missing data.
\end{itemize} 
Note that defining what must be covered (i.e., assuring the comprehensiveness property) and how to make the descriptions accurate (i.e., assuring the accuracy property) are two open problems on their own, which are orthogonal to the focus of the present study, which addresses the missing-data problem. 
Unfortunately, real-world cyber attack datasets often miss detailed technical descriptions that are required to properly understand the cyber attacks, as evidenced by the specific dataset we will analyze.

\subsubsection{Collecting Raw Data} 
Real-world space cyber attack data is often buried in raw CTI (Cyber Threat Intelligence) reports in the Internet, typically in the form of non-technical news reports as well as blog and social media posts. This means such data can be collected via Internet search, using keywords such as ``space AND cyber AND incident OR \{system, infrastructure, satellite, attack, interference, eavesdropping, spoofing, weapon, jamming, hijacking, seizure, corruption, interception, denial, deception\}.'' 
Given that there can be many ways to collect such data, in this paper we assume that an analyst is given a set of raw CTI reports with claimed pertinence to space cyber attacks, bearing in mind that these reports may be low quality and could be incorrect in the worst-case scenario.
The collected data is often 
in the form of unstructured narratives, and must go through a preprocessing step to assure relevance.




\subsubsection{Preprocessing Raw Data} \label{subsubsec:preprocessing}
This is to assure the relevance of each raw CTI report because some reports may be included by mistake, possibly incurred by errors of Internet search engines or incompetency of data sources. For this purpose, one approach is to leverage domain experts to filter the collected raw CTI reports according to the following attributes:
\begin{itemize}
    \item {\em Incident identification}: Each incident should be assigned a unique incident identifier
    for reference purposes, while avoiding duplicates.
    \item {\em Attack type}: 
    This refers to the attack type associated with an incident according to a chosen cyber risk taxonomy,
    such as the one presented in \cite{falco2021security}, which includes the following {\em attack types}: 
   High-powered Laser, High-powered Microwaves, RF Interferences, Eavesdropping, Spoofing, Ultrawideband Weapon, Electromagnetic Pulse (EMP) Weapon, Jamming, Signal Hijacking, Seizure of Control, Data Corruption/Interception, Denial of Service (DoS), and Space Situational Awareness (SSA) Deception. 
    \item {\em Date}: This is the date 
    when an incident takes place, ideally including 
    a unified time zone (e.g., the Greenwich Mean Time or GMT) if available. 
    \item {\em Geographic locations}: This describes where an incident takes place, ideally including 
    facility, city, state, and country. 
    The presence of such location information is an indicator of incident credibility.
    \item {\em Attack description}: 
    This is the description of an incident; the more detailed the better. There is often unstructured text in a raw CTI report, which
    requires substantial manual verification to ensure the description makes technical sense.
    \item {\em Attacker identity}: This is the name, alias, and/or country of origin of the attacker that incurred an incident. The attacker can be an individual, organization, company, or agency.  
    \item {\em Victim identity}: This is the name and industry of the victim involved in an incident. 
    This is another indicator of the credibility of the raw CTI reports.
    \item {\em Data sources}: These are the sources that report an incident. Multiple independent sources reporting the same incident indicate a higher incident credibility.
\end{itemize}
The result is a preprocessed dataset of unique space cyber attack incidents, where every incident is represented by a row of the attributes defined above.
In the ideal case, the preprocessed dataset is {\em comprehensive},
{\em accurate}, and containing {\em zero missing data} as described above.

\subsubsection{Processing Preprocessed Data with Full Information}
The task is to construct a $\USCKC$ from the description of a space cyber attack, while leveraging the afore-reviewed cyber threat taxonomies (i.e., SPARTA \cite{SPARTA}, 
ATT\&CK
\cite{ATTCK}, and UKC \cite{pols2017unified}). We reiterate the choice of these taxonomies as follows: (i) they are widely used by practitioners, making our results easily adoptable by them;  
and (ii) they collectively allow us to create USCKCs across the 4 segments because ATT\&CK focuses on enterprise IT networks in the ground segment, SPARTA focuses on the space segment, and UKC provides information for sequencing cyber attack progressions.
Note 
that our framework can be adapted to accommodate other cyber threat taxonomies (e.g., the ones that may be introduced in the future). 
However, SPARTA, ATT\&CK, and UKC do not provide any explicit method that can be leveraged to compile attack tactics or techniques into a $\USCKC$. This prompts us to propose a {\em skeleton} algorithm, Algorithm \ref{alg:processing_preprocessed_data},
to construct a $\USCKC$ for each preprocessed space cyber attack description with full information (i.e., with no missing data), where {\em skeleton} means that the algorithm is extracted from our manual processing and can guide the design of automated algorithm in the future. 

\begin{algorithm}[!htbp]
\label{alg:processing_preprocessed_data}
\caption{Constructing a $\USCKC$ of a space cyber attack with no missing data}
\KwInput{
attack description $\AttackDescription$ of the attack; 
space infrastructure graph $G = (V,E)$;
the joint SPARTA-ATT\&CK attack tactics set $\AvailableCapabilitySet_{\TA}$; the joint SPARTA-ATT\&CK attack techniques set $\AvailableCapabilitySet_{\TE}$}

\KwOutput{$\USCKC=\{\PH, \AC, \TA,\TE\}$, 
where $\PH = \{ph_1,\ldots,ph_n\}$ is the ordered set of attack phase types, 
${\AC} = \{ac_1,\ldots,ac_n\}$ is the ordered set of attack activity types, 
${\TA} = \{ta_1,\ldots,ta_n\}$ is the ordered set of attack tactics, and 
$\TE=\{te_1,\ldots,te_n\}$ is the ordered set of attack techniques that are employed by the attack} 

$\PH \gets\{\}$, $\AC \gets\{\}$, $\TA \gets\{\}$, $\TE \gets\{\}$ \tcp{initializing empty ordered sets}

partition $\AttackDescription$ into $n$ attack steps in order


\For{$i=1 ~\text{to}~n$}
{
determine attack phase $ph_i$ to which step $i$ belong and append $ph_i$ to $\PH$

determine activity type $ac_i$ to which step $i$ belong and append $ac_i$ to $\AC$

determine attack tactics $ta_i$ to which step $i$ belong and append $ta_i$ to $\TA$

determine attack technique $te_i$ to which step $i$ belong and append $te_i$ to $\TE$

}



$\USCKC\gets\{\PH,\AC,\TA,\TE\}$

\Return{$\USCKC$}

\end{algorithm}

At a high level, Algorithm \ref{alg:processing_preprocessed_data} generates one $\USCKC$, including both tactic-level and technique-level manifestations, per space cyber attack in the preprocessed
dataset. (For any entry in the preprocessed dataset with partial information or missing data, we will address it in Section \ref{sec:addressingMissingData}).
Its input includes the description $\AttackDescription$ of a space cyber attack, the context space infrastructure $G$, the set of joint SPARTA-ATT\&CK  attack tactics $\AvailableCapabilitySet_{\TA}$ and the set joint SPARTA-ATT\&CK  attack techniques $\AvailableCapabilitySet_{\TE}$ that are defined by desired versions of SPARTA and ATT\&CK.
Its output includes $\USCKC=\{\PH, \AC, \TA,\TE\}$ of $n$ attack steps,
where ${\PH}=\{ph_1,\ldots,ph_n\}$ is an ordered set of {\em phase types} with $ph_i\in \{\In, \Through, \Out\}$ for $1\leq i \leq n$ being the phase type to which the $i$th attack step belong, 
${\AC}=\{ac_1,\ldots,ac_n\}$ is the ordered set of {\em activity types} with $ac_i\in \{{\rm objective},{\rm milestone},{\rm enabling},{\rm  information~discovery}\}$ being the activity type to which the $i$th attack step belong, 
${\TA}=\{ta_1,\ldots,ta_n\}$ is the ordered set of {\em attack tactics} with $ta_i \in \AvailableCapabilitySet_{\TA}$ being the attack tactic to which the $i$th attack step belong,
and 
$\TE=\{te_1,\ldots,te_n\}$ is the ordered set of {\em attack techniques} with $te_i \in \AvailableCapabilitySet_{\TE}$ being the attack technique that is used in the $i$th attack step.

More specifically, the algorithm has 2 stages, {\em partitioning} (Line 2) and {\em classifying} (Lines 3-7), as follows.
Specifically, Line 1 initializes the ordered empty sets $\PH$, $\AC$, $\TA$, and $\TE$. Line 2 partitions $\AttackDescription$ into $n$ attack steps. In loop iteration for the $i$th attack step, Lines 4-7 determine the corresponding attack phase $ph_i\in \{\In,\Through,\Out\}$, activity type $ac_i \in \{{\rm objective}, ~{\rm milestone}, ~{\rm enabling},~{\rm information~ discovery}\}$, attack tactic $ta_i\in \AvailableCapabilitySet_{\TA}$, and attack technique $te_1\in \AvailableCapabilitySet_{\TE}$. 
We note that it is non-trivial to automate these steps (i.e., Lines 2 and 4-7), even leveraging Large Language Models. This is because the algorithm requires, for instance, a deep understanding of $G=(V,E)$
and $\AvailableCapabilitySet_{\TE}$ in order to distinguish an attack phase $ph_i=\In$, meaning the attacker compromises a module $v \in V$ as an {\em entry node} (typically from outside the space infrastructure), from an attack phase $ph_i=\Through$, meaning the attacker cannot directly compromise $v$ from outside the space infrastructure.
Correspondingly, the computational complexity of the algorithm depends on how each function is implemented; in the present study, they are realized manually and thus we cannot report the complexity.

\subsubsection{Addressing the Missing-Data Problem} \label{sec:addressingMissingData}


In practice, we often encounter missing data in the raw and preprocessed datasets, especially the attack description $\AttackDescription$.
In this case, Algorithm \ref{alg:processing_preprocessed_data}, which requires full information, cannot be used.
To address this missing-data problem, we propose extrapolating the missing data, which is a non-trivial task that demands a deep understanding of space infrastructure $G$, the joint attack tactics set $\AvailableCapabilitySet_{\TA}$, and the joint attack techniques set $\AvailableCapabilitySet_{\TE}$.
For example, a proper understanding of $G$ is necessary to determine whether a $\USCKC$ requires no $\Through$ phases, which can happen when the attacker's {\em entry node} and {\em objective node}
are the same, or multiple $\Through$ phases, which can happen when the attacker's entry node and objective node belong to different modules, components or even segments; 
a deep understanding of $\AvailableCapabilitySet_{\TA}$ is required 
to determine if the identified 
attack tactics can work together to achieve the attacker's overall objective; and
an in-depth understanding of $\AvailableCapabilitySet_{\TE}$ is required to identify when $te_i$ requires $te_{i-1}$ as a prerequisite or precondition. These explain why we have not been able to automate this algorithm and why we achieve this manually in the present study, while noting that it is non-trivial to apply Large Language Models to tackle the problem.

To address this missing-data problem,
we propose a {\em skeleton} Algorithm \ref{alg:data extrapolation} to construct a set of  
probable $\USCKC$s for each $\AttackDescription$.
Its input includes the context space infrastructure $G$, the set of joint attack tactics $\AvailableCapabilitySet_{\TA}$ and the set joint attack techniques $\AvailableCapabilitySet_{\TE}$ that are defined by desired versions of SPARTA and ATT\&CK.
Its output includes $\{\USCKC\}$ where
$\USCKC=\{\PH, \AC, \TA,\TE\}$ of $s$ attack steps with $s$ to be determined (unlike Algorithm \ref{alg:processing_preprocessed_data}),
${\PH}=\{ph_1,\ldots,ph_s\}$ is an ordered set of {\em phase types} with $ph_i\in \{\In, \Through, \Out\}$ for $1\leq i \leq s$ being the phase type to which the $i$th attack step belong, 
${\AC}=\{ac_1,\ldots,ac_s\}$ is the ordered set of {\em activity types} with $ac_i\in \{{\rm objective},{\rm milestone},{\rm enabling},{\rm  information~discovery}\}$ being the activity type to which the $i$th attack step belong, 
${\TA}=\{ta_1,\ldots,ta_s\}$ is the ordered set of {\em attack tactics} with $ta_i \in \AvailableCapabilitySet_{\TA}$ being the attack tactic to which the $i$th attack step belong, and 
$\TE=\{te_1,\ldots,te_s\}$ is the ordered set of {\em attack techniques} with $te_{i} \in \AvailableCapabilitySet_{\TE}$ being the attack technique that is used in the $i$th observed or extrapolated attack step.

More specifically, Algorithm \ref{alg:data extrapolation} proceeds in 3 stages, {\em partitioning} (Line 2), {\em extrapolating} (Lines 3-15), and {\em combining} (Line 16), as follows.
Line 1 initializes ordered empty sets $\PH$, $\AC$, $\TA$, and $\TE$. 
Line 2 partitions $\AttackDescription$ into $n'$ {\em observed} attack steps, while noting that there may be {\em hidden} or {\em unobserved} steps.
For the $i$th observed attack step ($1\leq i \leq n'$), 
Lines 4-11 obtain the observed attack step, denoted by subscript $(i,1)$ and extrapolate the 
$|J_i|$ missing attack steps prior to the observed attack step. 
The total number of steps is $s=\sum_{i=1}^{n'} (1-J_i)$ because an observed attack step $i$ is extrapolated to $(1-J_i)$ steps (including the observed step), while noting that $J_i\leq 0$. 
By renaming the sequence $(X_{1,J_1+1}, \ldots, X_{1,1}, X_{2,J_2+1}, \ldots, X_{2,1},\ldots, X_{n',J_{n'}+1},\\ \ldots, X_{n',1})$ as $(X_1,\ldots,X_s)$, where $X\in\{ph,ac,ta,te\}$, and by renaming $(k_{1,J_1+1}, \ldots, k_{1,1}, k_{2,J_2+1}, \ldots, k_{2,1},\ldots, k_{n',J_{n'}+1}, $ $\ldots, k_{n',1})$ as $(k_1,\ldots,k_s)$, 
Line 16 generates up to $K=\prod_{z=1}^s k_s$ $\USCKC$'s, where each $\USCKC$ is similar to what is generated in Algorithm \ref{alg:processing_preprocessed_data} but with length $s$.
Note that in our manual approach for Line 16, we may eliminate some extrapolated attack steps to make better space cybersecurity sense for a particular $\USCKC$, reducing the number of attack step from $s$ to $s'$, where $s' < s$. 
Similar to Algorithm \ref{alg:processing_preprocessed_data}, the computational complexity of the algorithm depends on how each function is implemented; in the present study, they are realized manually and thus we cannot report the complexity.

\begin{algorithm}[!htbp]
\label{alg:data extrapolation}
\caption{Constructing $\USCKC$s of space cyber attacks with missing data}
\KwInput{description $\AttackDescription$ of an attack with missing data;
space infrastructure graph $G = (V,E)$;
the joint SPARTA-ATT\&CK attack tactics set $\AvailableCapabilitySet_{\TA}$; the joint SPARTA-ATT\&CK attack techniques set $\AvailableCapabilitySet_{\TE}$}

\KwOutput{a set $\{\USCKC\}$ of probable $\USCKC$s extrapolated from $\AttackDescription$, where $\USCKC=\{\PH, \AC, \TA,\TE\}$}

$\PH \gets\{\}$, $\AC \gets\{\}$, $\TA \gets\{\}$, $\TE \gets\{\}$ \tcp{initializing empty ordered sets}

partition $\AttackDescription$ into $n'$ attack steps of attack techniques, in sequential order $1, 2,\ldots$ \tcp{the $n'$ steps are given or observed}

\For{$i=1 ~\text{to}~n'$}
{
    $j_i \gets 1$ \tcp{keeping extrapolated steps}

\ignore{
    \If{step $i$ does not require a prior attack step}{
    $J_i\gets J_i+1$
        
    determine attack phase $ph_j$ to which attack step $j$ belong and append $ph_x$ to $\PH$
        
    determine activity phase $ac_j$ to which attack step $j$ belong and append $ac$ to $\AC$
        
    determine attack tactics $ta$ to which attack step $j$ belong and append $ta$ to $\TA$
        
    determine attack technique $te$ to which attack step $j$ belong and append $te$ to $\TE$
}
} 
    \Repeat{no more prior steps need to be extrapolated}{

    determine attack phase $ph_{i,j_i}$ to which the current attack step ${(i,j_i)}$ belong 
        
    determine activity type $ac_{i,j_i}$ to which the current attack step ${(i,j_i)}$ 
        
    determine attack tactics $ta_{i,j_i}$ to which  the current attack step $(i,j_i)$ belong 
        
    determine a set of $k_{i,j_i}$ probable techniques $\TE_{i,j_i}=\{te_{i,j_i,1},\ldots,te_{i,j_i,k_{i,j_i}}\}$ that can be used at the present attack step $(i,j_i)$ \tcp{where $|\TE_{i,1}|=1$}

    $j_i\gets j_i-1$

    \ignore{
        extrapolate the prior attack step $j_i$ with multiple variations as required
            
        determine attack phases $\{ph_x\}$ to which step $x$ belong and append $\{ph_x\}$ to $\PH$
        
        determine activity phase $\{ac_x\}$ to which step $x$ belong 
        
        determine attack tactics $\{ta_x\}$ to which step $x$ belong 
        
        determine a set of probable attack techniques $\{\{te_x\}\}_{i,j_i}$ to which step $x$ belong and append $\{te_x\}$ to $\TE$
     }       
    }
    $J_i\gets j_i$ \tcp{bookkeeping, $J_i\leq 0$}
    
    $\PH\gets \{ph_{i,J_i+1},ph_{i,J_i+2},\ldots,ph_{i,1}\}$

    $\AC\gets \{ac_{i,J_i+1},ac_{i,J_i+2},\ldots,ac_{i,1}\}$
    
    $\TA\gets \{ta_{i,J_i+1},ta_{i,J_i+2},\ldots,ta_{i,1}\}$

}
    
$\{\USCKC\}\gets\{\PH,\AC,\TA,\TE\}$ where     $\TE= \TE_{1,J_i+1} \times \ldots \times \TE_{1,1}\times \ldots \times \TE_{n',J_i+1} \times \ldots \times \TE_{n',1}$ and each $\USCKC$ makes space cybersecurity sense

\Return{$\{\USCKC\}$}

\end{algorithm}

\subsection{Defining Metrics to Characterizing Attacks}

We define three metrics to characterize cyber attacks against space infrastructures: {\em attack consequence}, {\em attack sophistication}, and {\em $\USCKC$ Likelihood}, each of which is a multi-dimensional vector that can be aggregated into a single number if desired. These metrics are useful because (e.g.) one can use the first one to compare multiple attacks in terms of their consequences or damages and the second and to characterize how advanced an attack is.

\subsubsection{Attack Consequence Metric}    
    
Since space infrastructures have 4 segments and attack consequences may be manifested at some or all of the 4 segments, we define a vector of vectors, denoted by $(\vec{s}_\S,\vec{g}_\G,\vec{u}_\U,\vec{l}_\L)$, to represent attack consequences to the space segment (\S), ground segment (\G), user segment (\U), and link segment (\L), respectively. Elaborations follow.

\smallskip

\noindent{\bf Consequence to space segment ($\vec{s}_\S$)}.
We define the attack consequence to a Space Segment $\S$ as $\vec{s}_\S=(\vec{s}_\BS,\vec{s}_\PL)$, where vector $\vec{s}_\BS$ denotes the consequence to the Bus System (\BS) and vector $\vec{s}_\PL$ denotes the consequence to the Payload ($\PL$). 
\begin{itemize}
\item We define $\vec{s}_\BS=(s_{\BS,1},\ldots,s_{\BS,6})$ as a vector of attack consequences to the 6 components of the Bus System: electrical power ($s_{\BS,1}$), attitude control ($s_{\BS,2}$), communication ($s_{\BS,3}$), command \& data ($s_{\BS,4}$), propulsion ($s_{\BS,5}$), and thermal control ($s_{\BS,6}$). We define $s_{\BS,j}\in [0,1]$, $1\leq j \leq 6$, as the degree of the functionality (i.e., availability) 
of the corresponding component being degraded because of the attack in question, where $s_{\BS,j}=0$ (or 1) means the functionality is 0\% (or 100\%) degraded. 

\item We define $\vec{s}_\PL=(s_{\PL,1},\ldots,s_{\PL,5})$ as a vector of attack consequences to the 5 payload components: communication ($s_{\PL,1}$), navigation ($s_{\PL,2}$), scientific application ($s_{\PL,3}$), remote sensing ($s_{\PL,4}$), and national security ($s_{\PL,5}$), with $s_{\PL,j}\in [0,1]$ in the same fashion as $s_{\BS,j}$.

\end{itemize}
The preceding definition of $\vec{s}_\S$ has several salient features, which also apply to the subsequent metrics corresponding to the other segments.
First, we differentiate ``defining what to measure'' from ``how to measure what we need to measure.'' 
The present study addresses the former. Note also that our definitions remain valid when considering the inter-dependencies between different segments. Concerning the latter, it would be ideal that measurements of these metrics are provided for purposes such as the present study;
otherwise, one may use its domain expertise to estimate these metrics.
Second, we make the number of components specific to the system model described in Figure \ref{fig:system_model} (e.g., 6 components in the Bus System) to make the definitions easier to follow. The definitions can be trivially generalized to accommodate an arbitrary number of components. 
Third, the ``fine-granularity'' of $\vec{s}_\BS$ and $\vec{s}_\PL$ makes them suitable to compare consequences of multiple attacks and to make statements like ``Attack 1 is more powerful than Attack 2.''
Fourth, 
one can aggregate the vector metrics 
into a single number. For example,  $\vec{s}_\BS$ can be aggregated into $\bar{s}_\BS$ via some mathematical function $f$, namely  $\bar{s}_\BS=f(s_{\BS,1},\ldots,s_{\BS,6})$, where $f$ can be the (weighted) algebraic average function,
which makes cybersecurity sense because they deal with the same property (i.e., availability).
We can similarly aggregate $\bar{s}_\BS$ and $\bar{s}_\PL$ 
into a single number.

\ignore{
We may need to aggregate $v\vec{s}_P=(s_{P,1},\ldots,s_{P,5})$ into a single number, denoted by $\bar{s}_P$, as some function $f_{s_P}$, namely  $\bar{s}_P=f_{s_P}(s_{P,1},\ldots,s_{P,5})$. For example, one specific instance is the algebraic average, meaning $\bar{s}_P=\frac{1}{5} \sum_{k=1}^5 s_{P,k}$. One immediate extension is to consider the varying importance of the component. Suppose the weighted importance of the components are respectively $w_{P,1},\ldots,w_{P,5}$ where $0\leq w_{P,k} \leq 1$ and $\sum_{k=1}^5 w_{P,k}=1$. Then, weighted average of consequence on the Payload Applications incurred by an attack can be computed as $\bar{s}_P=\frac{1}{5} \sum_{k=1}^5 w_{P,k} \times s_{P,k}$.
}

\smallskip

\noindent{\bf Consequence to ground segment ($\vec{g}_\G$)}.
We define the attack consequence against a Ground Segment $\G$ as $\vec{g}_\G=(\vec{g}_{\GS},\vec{g}_{\MC},$ $\vec{g}_{\DPC},\vec{g}_{\RT})$, where $\vec{g}_{\GS}$ is the consequence to the Ground Station (\GS), $\vec{g}_{\MC}$ is the consequence to Mission Control (\MC), $\vec{g}_{\DPC}$ is the consequence to the Data Processing Center (\DPC), $\vec{g}_{\RT}$ is the consequence to the Remote Terminal (\RT). 
\begin{itemize}
\item We define $\vec{g}_{\GS}=(g_{\GS,1},\ldots,g_{\GS,4})$ as a vector of consequences to the 4 components of Ground Station: tracking ($g_{\GS,1}$), ranging ($g_{\GS,2}$), transmission ($g_{\GS,3}$), and reception  ($g_{\GS,4}$), where $g_{\GS,j}\in [0,1]$, $1\leq j \leq 4$, and 0 (1) means 0\% (100\%) functionality degradation. 

\ignore{

This ``fine-grained'' definition of $g_S$ can be used, for example, to compare the consequences incurred by two attacks so that we can make statement like ``attack 1 is more powerful than attack 2.''

We may need to aggregate $\vec{g}_S=(g_{S,1},\ldots,g_{S,4})$ into a single number, denoted by $\bar{g}_S$, as some function $f_{g_S}$, namely  $\bar{g}_S=f_{g_S}(g_{S,1},\ldots,g_{S,4})$. For example, one specific instance is the algebraic average, meaning $\bar{g}_S=\frac{1}{4} \sum_{j=1}^4 g_{S,j}$. One immediate extension is to consider the varying importance of the component. Suppose the weighted importance of the components are respectively $w_{S,1},\ldots,w_{S,4}$ where $0\leq w_{S,j} \leq 1$ and $\sum_{j=1}^4 w_{S,j}=1$. Then, weighted average of consequence on the Ground Station incurred by an attack can be computed as $\bar{g}_S=\frac{1}{4} \sum_{j=1}^4 w_{S,j} \times g_{S,j}$.

}

\item We define $\vec{g}_{\MC}=(g_{MC,1},g_{MC,2},g_{\MC,3})$ as a vector of consequences to the 3 components of the Mission Control: telemetry processing ($g_{\MC,1}$), commanding ($g_{\MC,2}$), and analysis and support ($g_{\MC,3}$), where $g_{\MC,j}\in [0,1]$, $1\leq j \leq 3$, is similar to $g_{\GS,j}$.


\ignore{

Similar to the definition of $g_S$, this method can be used to compare the consequences incurred by multiple attacks in terms of relative power.

We may need to aggregate $\vec{g}_M=(g_{M,1},\ldots,g_{M,3})$ into a single number, denoted by $\bar{g}_M$, as some function $f_{g_M}$, namely  $\bar{g}_M=f_{g_M}(g_{M,1},\ldots,g_{M,3})$. For example, one specific instance is the algebraic average, meaning $\bar{g}_M=\frac{1}{3} \sum_{k=1}^3 g_{M,k}$. One immediate extension is to consider the varying importance of the component. Suppose the weighted importance of the components are respectively $w_{M,1},\ldots,w_{M,3}$ where $0\leq w_{M,k} \leq 1$ and $\sum_{k=1}^3 w_{M,k}=1$. Then, weighted average of consequence on the Mission Control incurred by an attack can be computed as $\bar{g}_M=\frac{1}{3} \sum_{k=1}^3 w_{M,k} \times g_{M,k}$.

}

\item We define $\vec{g}_{\DPC}=(g_{\DPC,1},g_{\DPC,2})$ as a vector of consequence to the 2 components of Data Processing Center: mission analysis ($g_{\DPC,1}$) and payload processing ($g_{\DPC,2}$), where $g_{\DPC,j}\in [0,1]$ 
is similiar to $g_{\GS,j}$.

\ignore{

namely the degree of functionality of the component being degraded, where $g_{D,l}=0$ means the component is not affected by the attack in question, and $g_{D,l}=1$ means the component is completely degraded to a nonoperational status by the attack. Similar to the definition of $g_S$, this method can be used to compare the consequences incurred by multiple attacks in terms of relative power.

We may need to aggregate $\vec{g}_D=(g_{D,1},g_{D,2})$ into a single number, denoted by $\bar{g}_D$, as some function $f_{g_D}$, namely  $\bar{g}_D=f_{g_D}(g_{D,1},g_{D,2})$. For example, one specific instance is the algebraic average, meaning $\bar{g}_D=\frac{1}{2} \sum_{l=1}^2 g_{D,l}$. One immediate extension is to consider the varying importance of the component. Suppose the weighted importance of the components are respectively $w_{D,1},w_{D,2}$ where $0\leq w_{D,l} \leq 1$ and $\sum_{l=1}^2 w_{D,l}=1$. Then, weighted average of consequence on the Data Processing Center incurred by an attack can be computed as $\bar{g}_D=\frac{1}{2} \sum_{l=1}^2 w_{D,l} \times g_{D,l}$.

}

\item We define $\vec{g}_{\RT}=(g_{\RT,1},g_{\RT,2})$ as a vector of consequence to the components of the Remote Terminal: network access ($g_{\RT,1}$) and software access ($g_{\RT,2}$), where $g_{\RT,j}\in [0,1]$
as with $g_{\GS,j}$,

\ignore{

For $1\leq l \leq 2$, we define $g_{R,m}\in [0,1]$, namely the degree of functionality of the component being degraded, where $g_{R,m}=0$ means the component is not affected by the attack in question, and $g_{R,m}=1$ means the component is completely degraded to a nonoperational status by the attack. Similar to the definition of $g_S$, this method can be used to compare the consequences incurred by multiple attacks in terms of relative power.

We may need to aggregate $\vec{g}_R=(g_{R,1},g_{R,2})$ into a single number, denoted by $\bar{g}_R$, as some function $f_{g_R}$, namely  $\bar{g}_R=f_{g_D}(g_{R,1},g_{R,2})$. For example, one specific instance is the algebraic average, meaning $\bar{g}_R=\frac{1}{2} \sum_{m=1}^2 g_{R,m}$. One immediate extension is to consider the varying importance of the component. Suppose the weighted importance of the components are respectively $w_{R,1},w_{R,2}$ where $0\leq w_{R,m} \leq 1$ and $\sum_{m=1}^2 w_{R,m}=1$. Then, weighted average of consequence on the Remote Terminal incurred by an attack can be computed as $\bar{g}_R=\frac{1}{2} \sum_{m=1}^2 w_{R,m} \times g_{R,m}$.

}

\end{itemize}
Note that $\vec{g}_\G$ has the same salient features as $\vec{s}_\S$.

\smallskip

\noindent{\bf Consequence to user segment ($\vec{u}_\U$)}.
We define the attack consequence to User Segment $\U$ as $\vec{u}_\U=({u}_1,{u}_2,{u}_3)$, whose elements respectively measure the consequence to components of $\U$: transmission ($u_1$), reception  ($u_2$), and processing ($u_3$), with $u_{j}\in [0,1]$, $1\leq j \leq 3$, as with $s_{\BS,j}$. 
Note that $\vec{u}_\U$ has the same salient features as $\vec{s}_\S$ and $\vec{s}_\BS$.


\smallskip

\noindent{\bf Consequence to link segment ($\vec{l}_\L$)}.
We define the attack consequence to Link Segment as $\vec{l}_\L=(\{\vec{l}_{\S}\},\{\vec{l}_{\G}\},\{\vec{l}_{\SS}\}, \{\vec{l}_{\GG}\}, $ $\{\vec{l}_{\SG}\}, \{\vec{l}_{\SU}\}, \{\vec{l}_{\GU}\},\{\vec{l}_{\UU}\})$, where the elements correspond to a set of links within a space infrastructure
affected by an attack within a ground wide-area network (WAN),
between two satellites,
between two ground WANs,
between a Space Segment and a Ground Segment, between a Space Segment and a User Segment, between a Ground Segment and a User Segment, and between two users.
We further define: 
\begin{itemize}
\item $\vec{l}_\S=(l_{\S,\C},l_{\S,\I},l_{\S,\A})$ as the consequences to a link between the Bus System and the Payload, where $l_{\S,\C},l_{\S,\I},l_{\S,\A}\in [0,1]$ are the consequences to the confidentiality, integrity, and availability assurance of the link, with 0 (1) meaning 0\% (100\%) degradation. 

\item $\vec{l}_{\G}=(\vec{l}_{\GS,\MC},\vec{l}_{\GS,\DPC}, \vec{l}_{\GS,\RT},\vec{l}_{\MC,\DPC},
\vec{l}_{\MC,\RT},\vec{l}_{\DPC,\RT})$ as the consequences to links of components in a Ground Segment, where  $\vec{l}_{\GS,\MC}=({l}_{\GS,\MC;\C},{l}_{\GS,\MC;\I},{l}_{\GS,\MC;\A})$ $\in[0,1]^3$ are the consequences to the confidentiality, integrity, and availability of the link between a Ground Station (\GS) and a Mission Control (\MC), with 0 (1) meaning 0\% (100\%) degradation. 
The other elements of $\vec{l}_{\G}$ are defined in the same fashion.
 
\item $\vec{l}_{\SS}=(\vec{l}_{\SS,\C},\vec{l}_{\SS,\I},\vec{l}_{\SS,\A})$ as the consequences to a link between two satellites,
where $\vec{l}_{\SS,\C},\vec{l}_{\SS,\I},\vec{l}_{\SS,\A}\in[0,1]^3$ are the consequences to the confidentiality, integrity, and availability assurance of the link, with 0 (1) meaning 0\% (100\%) degradation.

\item $\vec{l}_{\GG}=(\vec{l}_{\GG,\C},\vec{l}_{\GG,\I},\vec{l}_{\GG,\A})$ as the consequences to a link between two ground WANs,
where $\vec{l}_{\GG,\C},\vec{l}_{\GG,\I},\vec{l}_{\GG,\A}\in[0,1]^3$ are the consequences to the confidentiality, integrity, and availability assurance of the link, with 0 (1) meaning 0\% (100\%) degradation.

\item $\vec{l}_{\SG}=(\vec{l}_{\SG,\C},\vec{l}_{\SG,\I},\vec{l}_{\SG,\A})$ as the consequences to a link between a Space Segment and a Ground Segment, where $\vec{l}_{\SG,\C},\vec{l}_{\SG,\I},\vec{l}_{\SG,\A}\in[0,1]^3$ are the consequences to the confidentiality, integrity, and availability assurance of the link, with 0 (1) meaning 0\% (100\%) degradation.

\item $\vec{l}_{\SU}=(\vec{l}_{\SU,\C},\vec{l}_{\SU,\I},\vec{l}_{\SU,\A})$ as the consequences to a link between a Space Segment and User Segment, where $\vec{l}_{\SU,\C},\vec{l}_{\SU,\I},\vec{l}_{\SU,\A}\in[0,1]^3$ are the consequences to the confidentiality, integrity, and availability assurance of the link, with 0 (1) meaning 0\% (100\%) degradation.

\item $\vec{l}_{\GU}=(\vec{l}_{\GU,\C},\vec{l}_{\GU,\I},\vec{l}_{\GU,\A})$ as the consequences to a link between a Ground Segment and User Segment, where $\vec{l}_{\SS,\C},\vec{l}_{\SS,\I},\vec{l}_{\SS,\A}\in[0,1]^3$ are the consequences to the confidentiality, integrity, and availability assurance of the link, with 0 (1) meaning 0\% (100\%) degradation.

\item $\vec{l}_{\UU}=(\vec{l}_{\UU,\C},\vec{l}_{\UU,\I},\vec{l}_{\UU,\A})$ as the consequences to a link between two users,
where $\vec{l}_{\UU,\C},\vec{l}_{\UU,\I},\vec{l}_{\UU,\A}\in[0,1]^3$ are the consequences to the confidentiality, integrity, and availability assurance of the link, with 0 (1) meaning 0\% (100\%) degradation.

\end{itemize}
Note that $\l_\L$ has the same salient features as $\vec{s}_\S$ except how metrics may be aggregated. For example, it would not make good cybersecurity sense to aggregate $(l_{\S,\C},l_{\S,\I},l_{\S,\A})$ via a (weighted) algebraic average because confidentiality, integrity, and availability metrics describe different properties.  One may suggest to aggregate them via $1-[(1-l_{\S,\C})(1-l_{\S,\I})(1-l_{\S,\A})]$, which is reminiscent of the aggregation of Common Vulnerability Scoring System (CVSS) scores  \cite{first2019common}. However, 
this does not appear sound as this aggregation function appears rooted in Probability Theory, which does not apply here because the events, even if $(l_{\S,\C},l_{\S,\I},l_{\S,\A})$ can be interpreted as probabilities, are not independent (e.g., the three assurances may be degraded at will by an attacker, rather than degrading independently of each other). Accordingly, we define:


\begin{definition}[Attack Consequence]
\label{definition:attack-consequence}
We define the attack consequence of a cyber attack against space infrastructures as 
$$\left(\cup_{\S} \{\vec{s}_\S\},~\cup_{\G} \{\vec{g}_\G\},~\cup_{\U} \{\vec{u}_\U\}, ~\cup_{\L}  \{\vec{l}_{\L} \}\right),$$
where the union is over all the Space Segments (${\S}$), Ground Segments (${\G}$), User Segments (${\U}$), and Link Segments (${\L}$) that are affected by the attack.
\end{definition}

\subsubsection{Attack Sophistication Metric}

We define this metric to characterize how sophisticated a cyber attack against a space system is through the lens of attack tactics and attack techniques.
Suppose for each cyber attack we extrapolate it into a set of $n$ $\USCKC$s, denoted by 
$\USCKC = \{\USCKC_1,\ldots,\USCKC_{n}\}$,
where $\USCKC_i$ ($1\leq i \leq n$) denotes the $i$th $\USCKC$ that is possibly associated with the cyber attack in question (when there is missing data describing the attack).
Suppose $\USCKC_i$ is associated with a set of $u$ attack {\em tactics}, denoted by $\TA_i=\{\TA_{i,1},\ldots,\TA_{i,u}\}$, and with a set of $v$ attack {\em techniques}, denoted by $\TE_i=\{\TE_{i,1},\ldots,\TE_{i,v}\}$,
where ${\text{TA}_{i,j}}$ and  ${\text{TE}_{i,j}}$ are respectively associated with a given sophistication score $\alpha_{\text{TA}_{i,j}}$ and $\alpha_{\text{TE}_{i,j}}$.
For $\USCKC_i$, we define its tactic sophistication as the maximum element among the element of $\TA_i$, namely $\max(\TA_i)=\max(\{\alpha_{\text{TA}_{i,1}},\ldots,\alpha_{\text{TA}_{i,u}}\})$, 
and define its attack technique sophistication as $\max(\TE_i)=\max(\{\alpha_{\text{TE}_{i,1}},\ldots,\alpha_{\text{TE}_{i,v}}\})$, which respectively corresponds to the most sophisticated attack tactic and attack technique used by the attacker. Now we are ready to define:

\begin{definition}[Attack Sophistication] 
\label{definition:attack-sophistication}
For an attack described by a set of hypothetical but plausible space cyber kill chains $\USCKC = \{\USCKC_1,\ldots,\USCKC_{n}\}$, we define its {\em possible highest sophistication} as vector $(\alpha_{\TA_+},\alpha_{\TE_+})$, where $\alpha_{\TA_+}=\max(\{\max(\TA_1),\ldots,\max(\TA_n)\})$, which corresponds to the most sophisticated attack tactic that is used among the possible kill chains, and $\alpha_{\TE_+}=\max(\{\max(\TE_1),\ldots,\max(\TE_n)\})$, which corresponds to the most sophisticated attack technique that is used among the possible kill chains.
We define its {\em possible lowest sophistication} as vector $(\alpha_{\TA_-},\alpha_{\TE_-})$, where $\alpha_{\TA_-}=\min(\{\max(\TA_1),\ldots,$ $\max(\TA_n)\})$, which corresponds to the least sophisticated attack tactic that is necessary for the attack to succeed, and $\alpha_{\TE_-}=\min(\{\max(\TE_1),\ldots,\max(\TE_n)\})$, which corresponds to the least sophisticated attack technique that is necessary for the attack to succeed.
\end{definition}

Note that Definition \ref{definition:attack-sophistication} can be easily extended to define the metric of, for example, {\em possible average sophistication}. 


\subsubsection{$\USCKC$ Likelihood} 
We define this metric to characterize the likelihood that a
$\USCKC$ may have been used to successfully accomplish a given space cyber attack. This is important because we often encounter missing details of space cyber attacks, which incur uncertainty on the $\USCKC$ that is actually used.

Suppose a 
$\USCKC$ has $n$ attack steps at the attack technique level of abstraction, namely one attack technique per attack step, denoted by an ordered set $\USCKC=\{te_1,\ldots, te_n\}$ where $te_i \in \TE$ and
$\USCKC$ may be obtained by Algorithm \ref{alg:data extrapolation}.
To compute the likelihood that a 
$\USCKC$ may have been used, namely $L(\USCKC_j)$, we assume a likelihood for each attack technique is given, either by domain experts or via a principled approach (the latter is an interesting future research problem).
We observe that $L(\USCKC_{j})$ is bounded by the lowest $L(te_i)$ because all attack techniques within the $\USCKC$ must succeed for the entire $\USCKC$ to succeed. 
In the case of $\{\USCKC\}$, only one $\USCKC \in \{\USCKC\}$ must succeed.
Hence, we define:

\begin{definition}[$\USCKC$ Likelihood] 
\label{definition:kill-chain-likelihood}
For a $\USCKC$ described at the technique level, namely $\USCKC=\{te_1,\ldots, te_n\}$, the likelihood that the $\USCKC$ succeeds is defined as $L(\USCKC) = \min(L(te_1), \ldots, L(te_n))$, where $L(te_\ell)$ for $1\leq \ell \leq n$ is the likelihood that attack technique $te_\ell$ is successful.
For a set $\{\USCKC\} = \{\USCKC_1,\ldots,\USCKC_m\}$, the likelihood that $\{\USCKC\}$ succeeds is defined as $L(\{\USCKC\}) = \max(L(\USCKC_1),\ldots,L(\USCKC_m))$.
\end{definition}

\subsection{Leveraging Metrics to Answer Research Questions}

Given a dataset of cyber attacks against space infrastructures, the metrics defined above can be leveraged to specify interesting research questions, such as:
(i) What is the attack consequence of each cyber attack against space infrastructures that has occurred in practice? Is there any trend exhibited by the attacks in terms of their attack consequence?
(ii) What is the attack sophistication of each cyber attack against space infrastructures that has occurred in practice? Is there any trend exhibited by the attacks in terms of their attack sophistication or attack entry points?
(iii) What is the uncertainty associated with the extrapolated $\USCKC$s? What is the likelihood of each extrapolated $\USCKC$?


\section{CASE STUDY}\label{sec:casestudy}

We apply the framework to characterize space cyber attacks based on a dataset of real-world space cyber incidents.


\subsection{Preparing Dataset}



\subsubsection{Collecting Raw Data}

In search of publicly available datasets, we first query ``satellite incidents,'' ``space attack data,'' and ``space incidents data'' in Google Scholar (with no limiting parameters), which return 45 papers.
Among these 45 papers, 3 papers \cite{pavur2022building, fritz2013satellite, falco2021security} are referred to by the other 42 papers, in terms of real-world space incidents that have affected space infrastructures; this means that these 42 papers do not provide any new incidents. 
The 42 papers also refer to an online database \cite{SpaceSecurityInfo} of space incidents. 
Specifically, 
70 incidents are reported in \cite{SpaceSecurityInfo}, 
112 in \cite{pavur2022building}, 
59 in \cite{fritz2013satellite}, and
1,847 in \cite{falco2021security}.We gathered the references provided in these four sources for each incident, leading to 272 raw CTI reports in total.
Then, another Google search with query ``satellite, space, incidents, attack, dataset'' returns many items, which include popular news articles and blogs, for which we reviewed the top 200 items.
From the 200 items, we identify 65 relevant items that
lead to 65 raw CTI reports.
This provides us a total of $272+65=337$ raw CTI reports.
We manually
review 
these 337 raw CTI reports, which are very diverse in terms of the number of incidents they contain. For instance, one CTI report found in \cite{falco2021security} actually contains 672 ``single event upset'' incidents where an electronic bit typically flips due to radiation in space, which are not about cyber attacks. Many 
other reported incidents found in the 337 CTI reports
are also not about cyber attacks,
such as incidents related to solar flare anomalies; these incidents are eliminated based on relevance. 
In total, we identify 162 space cyber attacks that make up our {\em raw dataset}.
Among these 162 space cyber attacks, 72
are reported in the conference version of the present paper \cite{ear2023CNS} and 90 are newly added in this version.

\subsubsection{Preprocessing Raw Data}

We use domain expertise to manually extract space cyber attack incidents from the 337 raw CTI reports resulting from the preceding step.
After eliminating duplicates,
we obtain a dataset of 108 unique space cyber attack incidents, among which: (i) 65 are extracted from three sources
\cite{SpaceSecurityInfo,pavur2022building,fritz2013satellite}, noting that these three sources substantially overlap with each other, 7 are extracted from \cite{falco2021security}, noting that these $65+7=72$ incidents are reported in the conference version
\cite{ear2023CNS};  
and (ii) 36 are extracted from the 90 incidents resulting from our Internet search mentioned above and are added in the present version.


The 108 attacks have different descriptions. Specifically, three datasets \cite{SpaceSecurityInfo,pavur2022building,falco2021security} describe attacks via: (i) year of incident occurrence; (ii) category of the attack/incident, such as jamming; (iii) attack target,
such as the USA-193 recon satellite; (iv) attacker identification,
such as nation-state; (v) narrative of facts concerning the incident, which are all brief; (vi) high-level attack objective, such as state-espionage;
and, (vii) source references, such as news articles reporting the incident.
On the other hand, the fourth dataset \cite{fritz2013satellite} only provides (i), (ii), (v), and (vii). Even so, they all lack some data.
Note that for (iv), the datasets provide attribution for 
76\% (86 out of the 108) space cyber attacks at varying degrees, ranging from country of origin to individual cyber threat organizations. 
The aforementioned 36 attacks are generally described in diverse ways in narrative form, where a few are lengthy case studies (e.g., of the ViaSat attack \cite{cyberpeace2022ViaSAT}), while the majority are short and lacking significant details.  


Consequently,
it is difficult to gauge the veracity of the 108 attacks.
For example, the RoSat incident is attributed to a
causation
between the cyber intrusion of the Goddard Space Flight Centre and the satellite's subsequent destructive maneuver in \cite{fritz2013satellite, Wess2021, Schneier2008cyberattacks, Epstein2008}, but this causation is refuted by a scientist on the RoSat project in an interview \cite{McDowell2011} and resonated by another source  \cite{soesanto2021terra}.
This indicates the {\em absence} of ground-truth attack details, meaning we must embrace the uncertainty imposed by missing-data while striving to leverage the available data to infer as much as possible.

Still, we manage to derive that the 108
space cyber attacks comprise 8 categories: 
26 Data Corruption/Interception attacks; 9 Denial of Service (DoS) attacks; 3 Eavesdropping attacks; 2 High-powered Laser attacks; 41 Jamming attacks; 3 Seizure of Control attacks; 15 Signal Hijacking attacks; and 9 Spoofing attacks. 
We also observe that 73\% of the attacks fall into three related categories: political, state espionage, and criminal; this indicates that space cyber attacks are likely state-sponsored.
We refer to the resulting dataset of $108$ space cyber attacks as the {\em 
preprocessed} space cyber attack dataset.

\subsubsection{Processing Preprocessed Data with Full Information}

Given that all the 108 attacks have missing data, Algorithm \ref{alg:processing_preprocessed_data} is not applicable to our case study. However, this does not mean the algorithm is useless. Instead, the algorithm is useful to researchers and practitioners who know the details of space cyber attacks (e.g., operators of space infrastructures).


\subsubsection{Addressing the Missing Data Problem}\label{sec:casestudy-addressingMissingData} 

As mentioned above, the description of all 108 attacks have missing data, which prompts us to
extrapolate attack phases, activities, tactics and techniques as described in Algorithm \ref{alg:data extrapolation}, as follows.
We manually partition the description of each of the 108 attacks into $n'$ observed or reported individual attack steps (Line 2), where $n'$ ranges from 1 to 9 steps, with average $\bar{n'}=2$. For example, the description of the RoSat attack contains the statement that the attacker caused the satellite to ``{\em miscalculate its alignment and turned toward the sun . . . NASA was able to correct}'' \cite{Wess2021}. For this attack, our manual analysis extracts $n'=9$ observed attack steps at the technique-level. 

For each of the $n'$ observed steps, we manually determine whether and how to extrapolate (Lines 3-15). 
Among all the observed steps and the 108 attacks, we extrapolate 
1 to 10 attack steps, meaning that we always extrapolate at least one step preceding a given attack step. Consider again the RoSat attack and the 5th (i.e., $i=5$) attack step of the observed $n'=9$ attack steps. 
This 5th attack step
requires network data to be successfully executed at the network access module, meaning the data must be provided by the previous step. This prompts us to extrapolate an attack step prior to the one in question, leading to attack phase $ph_{5,0} = \Through$, activity type $ac_{5,0}={\rm Information~Discovery}$, attack tactic $ta_{5,0}={\rm Reconnaissance}$; in terms of attack techniques, we hypothesize that the attacker requires network data to successfully bypass defense (i.e., to accomplish an {\em enabling} activity),  leading to $k_{5,0}=2$ probable attack techniques, namely $te_{5,0,1}=$ T1595 (Active Scanning ATT\&CK technique) and $te_{5,0,2}=$ T1590 (Gather Victim Network Information ATT\&CK technique).
We manually determine whether all the combinations obtained via the Cartesian product (Line 16) make space cybersecurity sense. 
\ignore{
Consider again the aforementioned RoSat attack. In total, our analysis leads to $s=14$ attack steps, including the $n'=9$ observed attack steps.
For example, the aforementioned RoSat attack prompts us to propose 2 extrapolated attack techniques for its {\color{olive}observed} Reconnaissance tactic step, namely $te_{5,0,1}=$ T1595  and $te_{5,0,2}=$ T1590 {\color{olive}corresponding to $s=14$ with $|k_{5,0}|=2$,} and 3 extrapolated attack techniques for its
9th {\color{olive}observed Persistence tactic step}, namely $te_{9,0,1}={\rm T1543}$ (Create or Modify System Process ATT\&CK technique), $te_{9,0,2}={\rm T1098}$ (Account Manipulation ATT\&CK technique), and $te_{9,0,3}={\rm T1133}$ (External Remote Services ATT\&CK technique) {\color{olive}corresponding to $s=14$ with $|k_{9,0}|=3$.} {\color{olive}These two steps alone already yield 6 combinations.}
}

\begin{figure*}[!htbp]
\centering
\includegraphics[width=1.8\columnwidth]{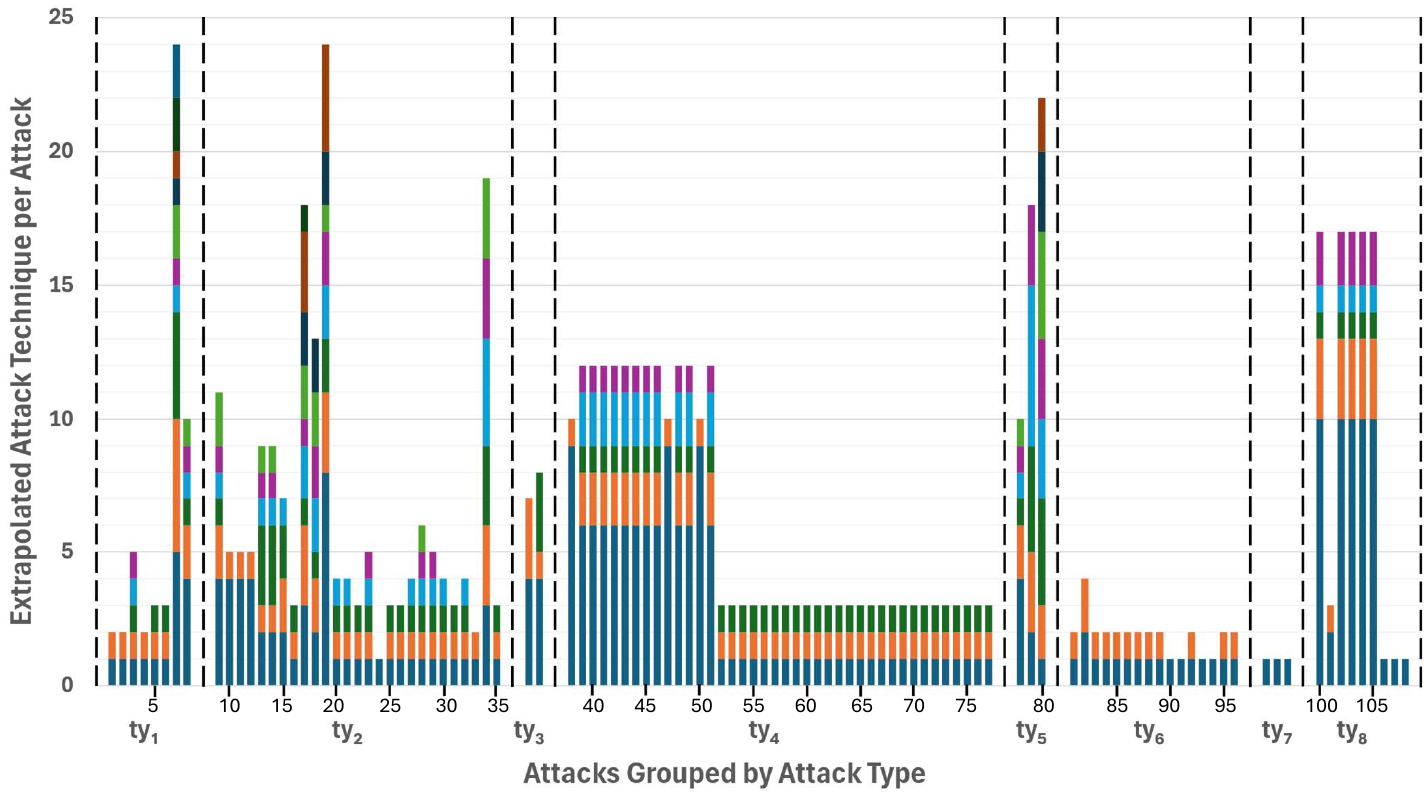}
\caption{The number of extrapolated attack techniques ($y$-axis) for each of the 108 attacks ($x$-axis). For each bar, the number of colors corresponds to the number of extrapolated attack steps, meaning that the length of a color corresponds to some $k_{i,j_i}$ in Algorithm \ref{alg:data extrapolation}; the product of the lengths of sub-bars in different colors is up to $|\{\USCKC\}|$ in Algorithm \ref{alg:data extrapolation}, namely the total number of probable $\USCKC$s extrapolated from an attack, because some $\USCKC$s may not make space cybersecurity sense. For instance, we extrapolated the RoSat attack (i.e., attack \#79) for 5 attack steps, leading to $2\times 3 \times 4 \times 6 \times 3 =432$ probable $\USCKC$s in total (i.e., they all make space cybersecurity sense).  
The 108 attacks are grouped according to their {\em attack type}, showing that 8 (out the aforementioned 13) attack types are observed, dubbed ty$_1$ (Denial of Service), ty$_2$ (Data Corruption/Interception), ty$_3$ (High-powered Laser), ty$_4$ (Jamming), ty$_5$ (Seizure of Control), ty$_6$ (Signal Hijacking), ty$_7$ (Eavesdropping), and ty$_8$ (Spoofing).}
\label{fig:extrapolation details}
\end{figure*}

Figure \ref{fig:extrapolation details}
plots the extrapolation results for the 108 attacks, where each bar corresponds to one attack and the attacks are organized according to the {\em attack type} to which they belong. 
For instance, the RoSat attack (\#79) leads to $n'=9$ and $s=14$, meaning 5 attack steps are extrapolated. For these 5 steps, 2, 3, 4, 6, and 3 attack steps are probable, leading to 
$2\times 3 \times 4 \times 6 \times 3 =432$ probable $\USCKC$s in total.
This large size of $\{\USCKC\}$ is indicative of the complexity of this attack and the large amount of missing data. We observe that attack types 1, 2, 5, and 8 have a larger number of probable $\USCKC$s, 
perhaps because they target the ground or user segment, where humans operate and interact with space infrastructures and offer attackers opportunities to wage cyber social engineering attacks, 
as opposed to the single entry node and deterministic behavior of the space segment or the physics-bound link segment. 
We further observe that attack types 1 and 3 target the ground segment, where attack capabilities effective against enterprise IT systems and networks 
are reusable against the ground segment (i.e., attackers can reuse their attack capabilities); attack type 5 is complex because these attackers often seeks to control the ground segment before seeking to control the space segment, as evidenced by the 3 attacks of this type in our dataset; and attack type 8 employs payloads that are often specific to their targets in the user segment.
In summary, we observe that the ground and user segments provide a larger attack surface than the space and link segments because they offer attackers opportunities to use more mature and a wider variety of attack capabilities.

\begin{insight}
There is a larger number of probable $\USCKC$s that target the ground and user segments than the ones that target the space and link segments perhaps because sophisticated attacks that directly target the space and link segments (without pivoting from the ground and user segments) are yet to be uncovered.
\end{insight}
We caveat our results with the consideration that ATT\&CK, which is geared towards the ground and user segments, is more mature than SPARTA, which is geared towards the space and link segments, contributing to the larger number of $\USCKC$s for attack types 1, 2, 5, and 8.

\begin{figure*}[!htbp]
\centering
\includegraphics[width=1.9\columnwidth]{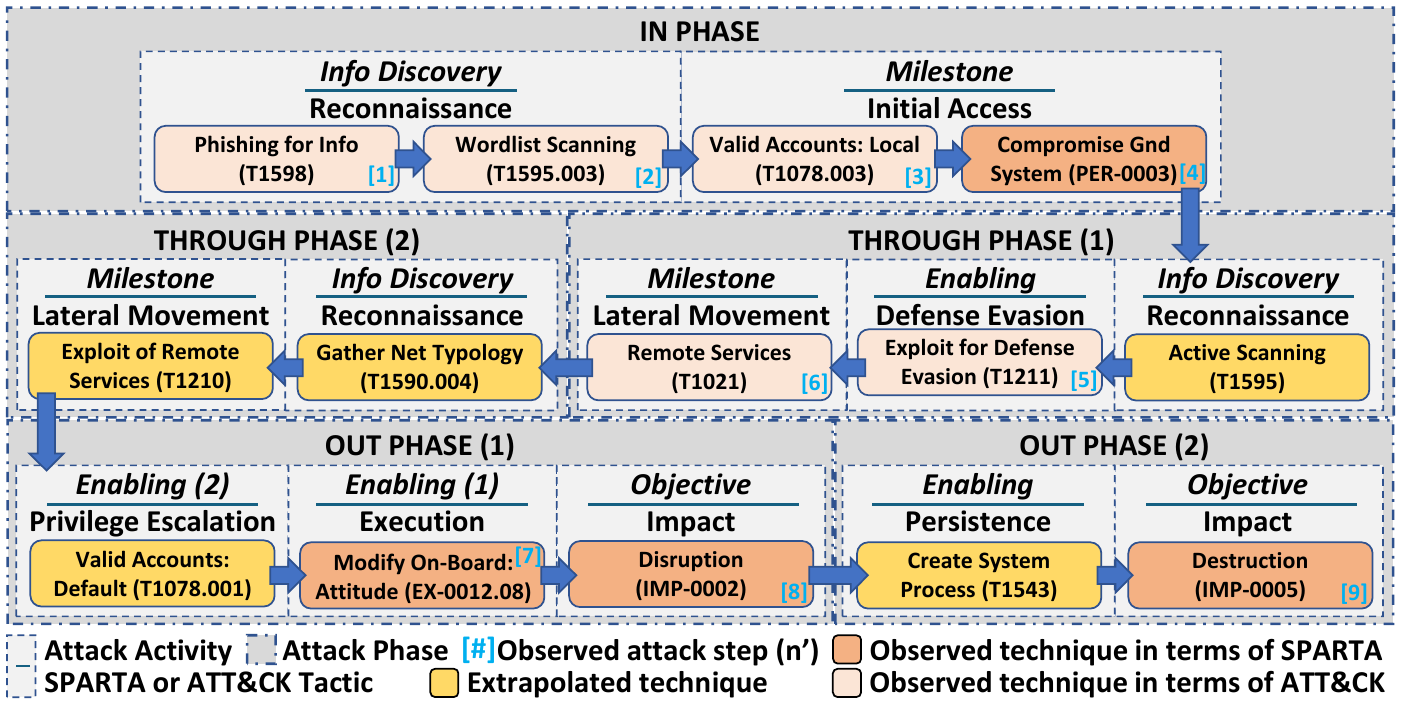}
\caption{One example $\USCKC$ (out of the 432 probable $\USCKC$s) we extrapolated for the RoSat 1998 attack, with $n'=9$ and $s=14$.}
\label{fig:example-kill-chain}
\end{figure*}

Figure \ref{fig:example-kill-chain} highlights one $\USCKC$ (out of 432) extrapolated from the RoSat attack. It can be understood as follows.
During the $\In$ phase,
the attacker uses techniques T1598 (Phishing for Information) and T1595.003 (Wordlist Scanning) to conduct a thorough reconnaissance, and uses techniques T1078.003 (Valid Accounts: Local) and PER-0003 (Compromise Ground System)  to gain initial access into the space infrastructure.
In the first $\Through$ phase,
the attacker uses an extrapolated technique T1595 (Active Scanning) to conduct further reconnaissance, uses technique T1211 (Exploit for Defense Evasion) to evades the defense, 
and uses technique T1021 (Remote Services) to support its 
lateral movement to get deeper into the space infrastructure.
During the second $\Through$ phase,
the attacker uses an extrapolated technique T1590.004 (Gather Network Typology) to accomplish further reconnaissance and then uses an extrapolated technique T1210 (Exploit of Remote Services) to laterally moves to a module that can affect the space segment. 
In the first $\Out$ phase,
the attacker uses an extrapolated technique T1078.001 (Valid Accounts: Default) to escalate its privileges, uses technique 
EX-0012.08 (Modify On-Board: Attitude) to execute its malware,
and uses technique IMP-0002 (Disruption) to temporarily impact the RoSat satellite. 
In the second $\Out$ phase, the attacker uses an extrapolated technique T1543 (Create System Process) to establish persistence in the space infrastructure and then uses technique IMP-0005 (Destruction) permanently to impact the RoSat's x-ray sensing device.

\begin{figure*}[!htbp]
\centering
\includegraphics[width=1.6\columnwidth]{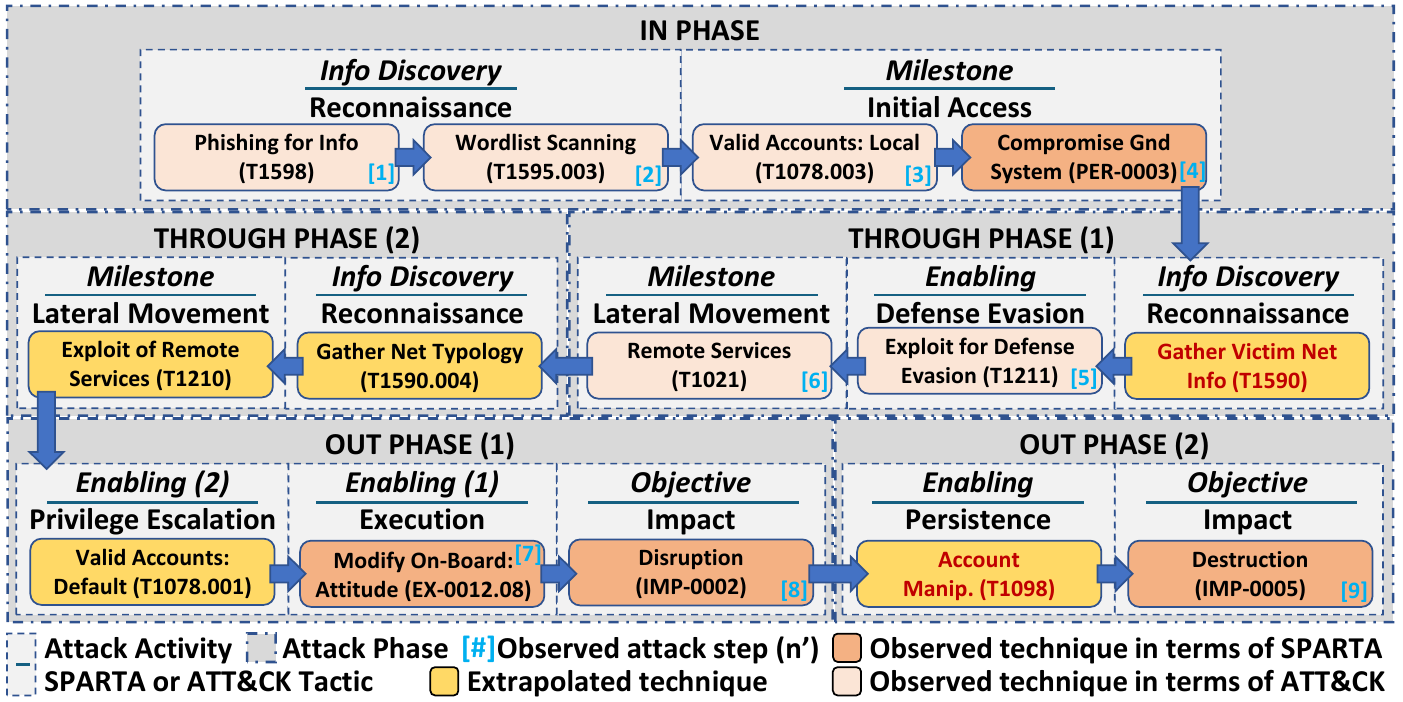}
\caption{Another example $\USCKC$ (out of the 432 hypothetical but plausible $\USCKC$s) we extrapolated for the RoSat 1998 attack.}
\label{fig:another example USCKC}
\end{figure*}

Figure \ref{fig:another example USCKC} presents another example of the 432 probable $\USCKC$s extracted from the RoSat attack. It can be understood as follows. 
During the $\In$ phase, the attacker uses techniques T1598 (Phishing for Information) and T1595.003 (Wordlist Scanning) to conduct reconnaissance and techniques 
T1078.003 (Valid Accounts: Local) and PER-0003 (Compromise Ground System)
to gain initial access to the space infrastructure.
In the first $\Through$ phase, the attacker uses an extrapolated technique T1590 (Gather Victim Network Information) to conduct reconnaissance, uses 
technique T1211 (Exploit for Defense Evasion) to evade the defense, and 
uses technique T1021 (Remote Services) 
to move laterally.
In the second $\Through$ phase, the attacker uses an extrapolated technique 
T1590.004 (Gather Network Typology) to conduct further reconnaissance and another extrapolated technique T1210 (Exploit of Remote Services) to support its lateral movement. 
In the first $\Out$ phase, the attacker uses an extrapolated technique T1078.001 (Valid Accounts: Default) to escalate its privileges, and uses technique EX-0012.08 (Modify On-Board: Attitude) to 
execute its malicious payload, and further uses technique IMP-0002 (Disruption) to temporarily impact the satellite.
In the second $\Out$ phase, the attacker uses an extrapolated technique T1098 (Account Manipulation) to achieve persistence, and uses technique IMP-0005 (Destruction)
to enable the attacker future access to impact the RoSat's x-ray sensing device.

The preceding two example $\USCKC$s are extrapolated from the same $n'=9$ observed attack steps. While differing in details, they both are probable  because: in the case of Figure \ref{fig:example-kill-chain}, the extrapolated T1595 ($te_{5,0, 1}$) provides the needed network data to inform T1211 ($te_{5,1}$), while the extrapolated technique T1543 ($te_{9,0, 1}$) provides the needed persistent access to enable IMP-0005 ($te_{9,1}$); in the case of Figure \ref{fig:another example USCKC}, T1590 ($te_{5,0, 2}$) provides the needed network data to inform T1211 ($te_{5,1}$), while T1098 ($te_{9,0, 2}$) provides the needed persistent access to enable IMP-0005 ($te_{9,1}$).

In total, we extrapolate 6,206 $\USCKC$s from the 108 attacks. We call it the {\em $\USCKC$ dataset}, where one row represents one $\USCKC$.
We will make this dataset available via an appropriate method.

\subsection{Leveraging Metrics to Answer Research Questions}


\ignore{

{\color{red}What trends have been exhibited by cyber attacks against space infrastructures in terms of attack impact, attack complexity, attack capabilities?}

...

Our resutls are limited in that our solution for RQ2 is specific to the particular dataset of this case study. Hence, our findings here may not be generalizable cross other datasets.

- high level insight: picture showing trend through the years
- number of attacks by type
{\color{purple}
The higher number of incidents that we have been able to compiled is 25 jamming incidents being followed by 18 CNE and 16 hijacking incidents. From that point on, the rest of categories only contain very few incidents, varying from 4 incidents to 1 incident.

}
    - incorporate origin of attacks

}

\subsubsection{Attack Consequence Analysis}


We leverage Definition~\ref{definition:attack-consequence}, our space infrastructure system model, our raw cyber attack dataset, and our extrapolated $\USCKC$ dataset to holistically consider the attack consequence via its manifestation in the space, ground, user, and link segments. For each attack, we leverage our domain expertise to assess the attack consequence in each segment, by assigning a score between 0 and 1, representing the least to most consequential. More specifically, we assign a score less than or equal to 0.3 when the attack impact is superficial (e.g., it is instantly recoverable); we assign a score between 0.3 and 0.8 when the attack impact is temporary (i.e., recoverable); we assign a score greater than or equal to equal to 0.8 when the attack impact is non-recoverable.

\begin{figure*}[!htbp]
\centering
\includegraphics[width=.9\textwidth]{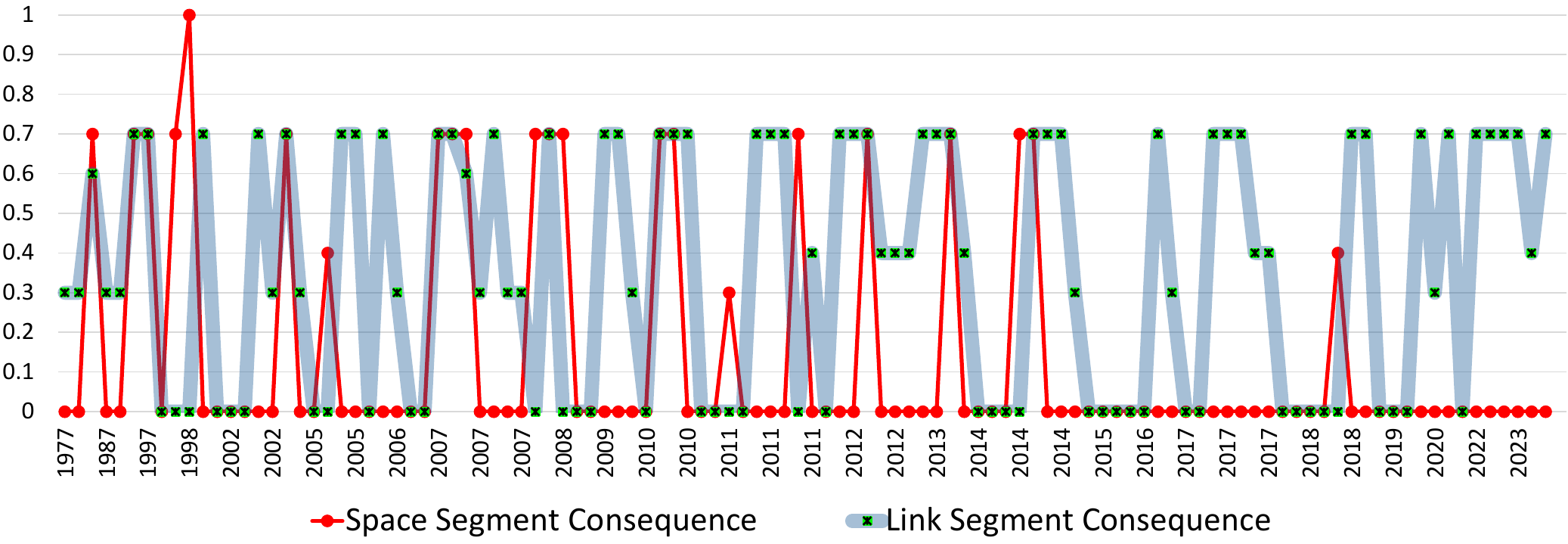}
\caption{Attack consequence in the space and link segments incurred by the 108 attacks over time.}
\label{fig:consequence-space-link}
\end{figure*}

Figure~\ref{fig:consequence-space-link}  shows that attack consequence for the space segment is generally low,
where less than 18\% of the attacks (i.e., 19 out of the 108) have a consequence score of 0.7 or higher: 1 Signal Hijacking, 12 Jamming, 3 DoS, and 3 Seizure of Control.
We observe that Seizure of Control attacks achieve the highest consequence against the space segment, where 2 attacks score 0.7 and one attack scores 1.0 (i.e., the 1998 RoSat incident scored the highest as it demonstrated the ability to physically destroy an asset in the space segment via cyber means). We observe that jamming is the most common attack type to produce a consequence in the space segment.

Figure~\ref{fig:consequence-space-link} also shows that attack consequence to the link segment has been consistent over the years. We observe 61\% of the attacks (i.e., 66 out of the 108) have a consequence score of 0.3 or higher, including 1 DoS attack, 41 Jamming attacks, 15 Signal Hijacking attacks, and 9 Spoofing attacks.
The 41 Jamming attacks have a high attack consequence score (0.7) because they degrade the conditions (e.g., noise level) of the channel employed by the legitimate signal, preventing the legitimate user and ground reception modules from receiving a proper signal. 
We also observe that Signal Hijacking and Spoofing attacks achieve lower consequence scores (i.e., 0.3 for Signal Hijacking and 0.4 for Spoofing) because they typically leverage signals to produce consequences to the other 3 segments.
The attacker in a Signal Hijacking or Spoofing attack cannot tolerate the downgrade of channel conditions, thus the consequence on the link segment is lower than what is incurred by jamming.



\begin{figure*}[!htbp]
\centering
\includegraphics[width=.9\textwidth]{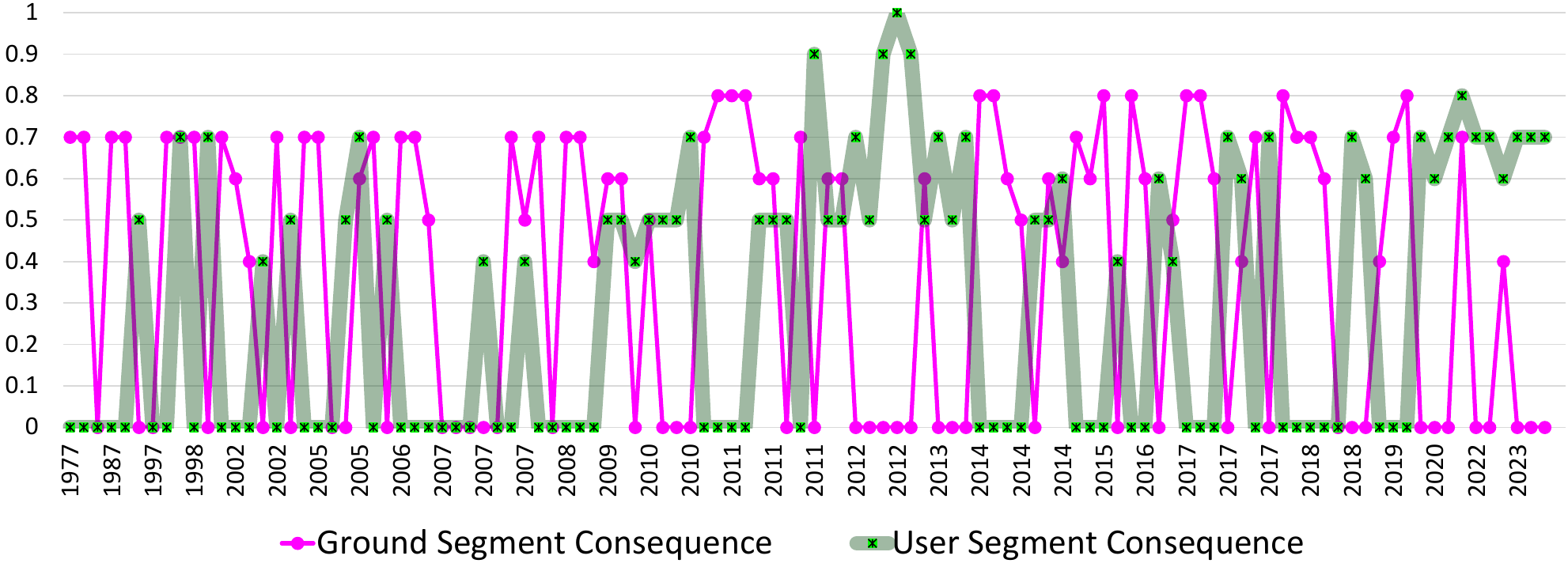}
\caption{Attack consequence in the ground and user segments incurred by the 108 attacks over time.}
\label{fig:consequence-ground-user}
\end{figure*}

Figure~\ref{fig:consequence-ground-user} shows that 
58\% of the attacks (i.e., 63 out of the 108) achieve a 
consequence of 0.4 or higher against the ground segment, including 26 Data Corruption/Interception attacks, 9 DoS attacks, 2 Eavesdropping attacks, 11 Jamming attack, 3 Seizure of Control attacks, 11 Signal Hijacking attacks, and 1 Spoofing attack,
while noting that the remaining 45 (out of the 108) attacks have no consequences in the ground segment.
We also observe that 11 (out of the 26) Data Corruption/Interception attacks achieve the highest consequence against the ground segment, with a consequence score of 0.8.
Figure~\ref{fig:consequence-ground-user} also shows that 49\% of the attacks (i.e., 53 out of the 108) have a consequence of 0.4 or higher against the user segment, including 4 DoS attacks, 1 Eavesdropping attack, 36 Jamming attack, 4 Signal Hijacking attacks, and 8 Spoofing attacks, while the remaining 55 (out of the 108) attacks have no consequences against the user segment. 
We also observe that 4 (out of the 8) Spoofing attacks achieve the hightest consequence against the ground segment,
including 1 Spoofing attack achieving a consequence of 1.0 (i.e., the University of Texas Radionavigation Laboratory demonstrated in 2012 the ability to make a rotorcraft dive to the ground by spoofing GPS signals).



\noindent{\bf How can we prioritize hardening 
to mitigate space cyber attacks?} 
We observe that 48\% (52 out of the 108) attacks achieve a consequence of 0.6 or higher against the ground segment, where 65\% (34 out of the 52) attacks leverage the ground segment as an entry point into the space infrastructure; the 34 attack include
24 Data Corruption/Interception attacks,
7 DoS attacks, and 3 Seizure of Control attacks. 
Further, the extrapolated $\USCKC$ dataset reveals victims exhibiting potentially weak security hygiene (e.g., default credentials) and unpatched public facing application vulnerabilities.
This leads to the following insight:

\begin{insight}
Attack consequences in the ground segment can be mitigated by prioritizing hardening measures in the ground segment (to prevent attackers from pivoting from the ground segment to the space segment).
\end{insight}

\subsubsection{Attack Sophistication Analysis}
\begin{figure*}[!htbp]
\centering
\includegraphics[width=1.8\columnwidth]{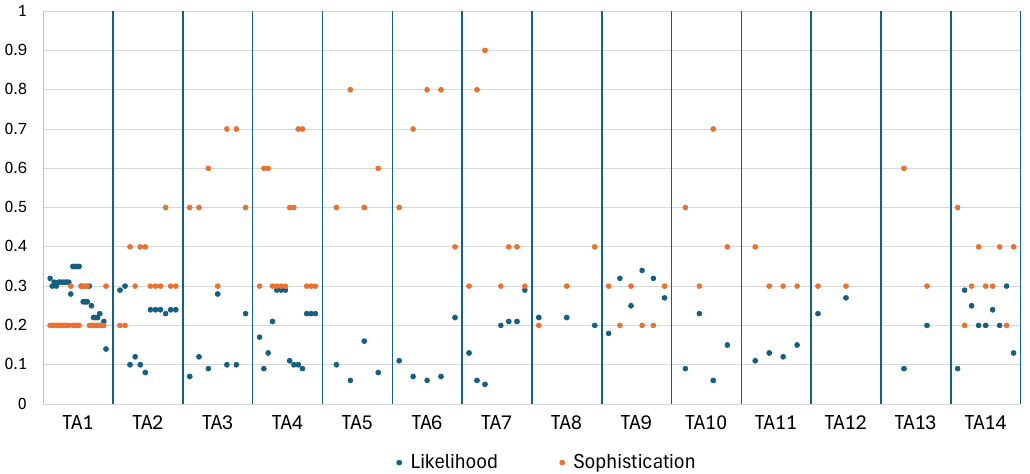}
\vspace{-1em}
\caption{Sophistication and likelihood measurements for the 107 attack techniques. TA1 = Reconnaissance; TA2 = Resource Development; TA3 = Initial Access; TA4 = Execution; TA5 = Persistence; TA6 = Privilege Escalation; TA7 = Defense Evasion; TA8 = Credential Access; TA9 = Discovery; TA10 = Lateral Movement; TA11 = Collection; TA12 = Command and Control; TA13 = Exfiltration; TA14 = Impact.}
\label{fig:sophistication-likelihood-per-at}
\end{figure*}

We use the sophistication metric (Definition \ref{definition:attack-sophistication}), namely $(\alpha_{\TA_+},\alpha_{\TE_+})$,  to quantify the {\em possible highest sophistication} for each attack according to its extrapolated attack tactic chain $\TA$ and its associated attack technique chains $\TE$'s, while taking as input the measurement of the sophistication score of each attack tactic and each attack technique, which are currently assigned based on our domain expertise and depicted in Figure \ref{fig:sophistication-likelihood-per-at}. Our results are as follows.

From the 108 attacks, we identify 14 attack tactics and 107 attack techniques in total.
For the 14 attack tactics, we manually score each from 0 to 1, considering 0.5 as the average sophistication required to accomplish an attack tactic. For example, we score the {\em Initial Access} tactic as 0.5 because majority of the 108
attacks should gain initial access to its target. We assign a score of 0.8 or higher for the tactics requiring a high sophistication. For example, the {\em Defense Evasion} tactic is scored at 0.9 as it requires additional effort and more advanced capabilities.
For the 107 attack techniques,
we assign sophistication scores while bearing in mind whether the score should be
less than, the same as, or greater than, 
that of the associated attack tactic. 
Phishing is commonplace and may require little technical capabilities when compared to the other Initial Access techniques, and hence receives a score of 0.3.
We then compute attack sophistication according to Definition \ref{definition:attack-sophistication}.



\begin{figure*}[!htbp]
\centering
\includegraphics[width=1.8\columnwidth]{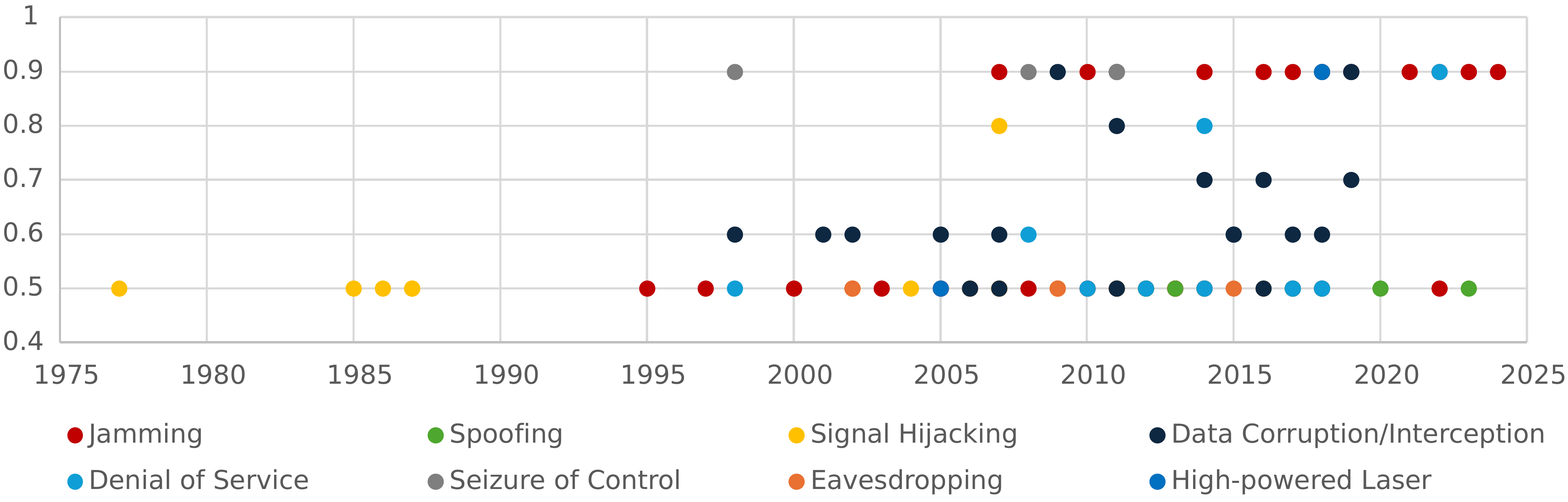}
\vspace{-1em}
\caption{Attack tactic sophistication of the 108 attacks over time. 
}
\label{fig:TA_soph_over_time}
\end{figure*}

Figure~\ref{fig:TA_soph_over_time} depicts the possible highest sophistication of the 108 attacks via $\alpha_{\TA_+}$, noting that several plots overlap and hide other plots from view. We observe that Signal Hijacking attacks consistently score 0.5, with 1 exception. In 2007, the Russian Turla Hacking Group's employment of the C2 tactic (ATT\&CK ID TA0011) by leveraging SATCOM connections scores 0.8, which we consider high sophistication. 
Approximately 70\% of the Jamming attacks score 0.5, and the remainder of Jamming attacks score 0.9 due to the inclusion of the Defense Evasion Tactic (ATT\&CK ID TA0005). 
Each Seizure of Control attack 
attains the high sophistication score of 0.9. Each of these attacks successfully employs the Defense Evasion tactic (ATT\&CK ID TA0005) to overcome ground control station defenses. Majority (19 out of the 26) of Data Corruption/Interception attacks score above 0.5, with one incident in 2011 scoring 0.9 because of its employment of both Defense Evasion and Persistence tactics to enable a series of 46 subsequent attacks against the ground segment. 
DoS attacks consistently score 0.5 except for 3 attacks that employ the Lateral Movement (ATT\&CK ID TA0008), Persistence (ATT\&CK ID TA0003), and Defense Evasion (ATT\&CK ID TA0005) tactics.
All High-powered Laser attacks in our dataset score 0.7 and all Eavesdropping and Spoofing attacks in our dataset score 0.5. Overall, space cyber attacks are getting more sophisticated over time.

\begin{figure*}[!htbp]
\centering
\includegraphics[width=1.8\columnwidth]{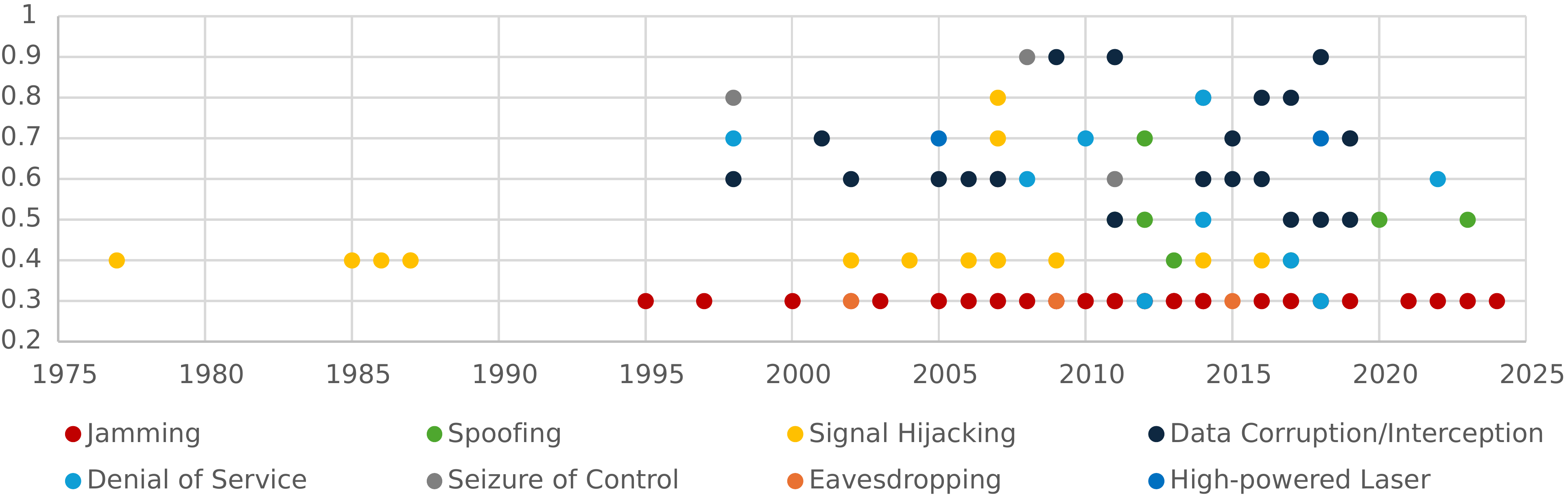}
\vspace{-1em}
\caption{Attack technique sophistication of the 108 attacks over time. 
}
\label{fig:TE_soph_over_time}
\end{figure*}

Figure~\ref{fig:TE_soph_over_time} depicts the possible highest sophistication of each attack via $\alpha_{\TE_+}$. We observe that $\alpha_{\TE_+}$ and $\alpha_{\TA_+}$ identify the same set of highly sophisticated space cyber attacks. The notable attack techniques employed by these attacks to support the Defense Evasion (T1211) tactic are: Indicator Removal (T1070) and Exploit (i.e., of a vulnerability).
The attacks that employ the Persistence tactic also employ the following attack techniques: Event Triggered Execution (T1546),  Create or Modify System Process (T1543), and Exploit Hardware/Firmware Corruption (EX-0005).

By comparing $\alpha_{\TA_+}$ and $\alpha_{\TE_+}$, we observe that these two metrics generally follow the same trend per incident,
but $\alpha_{\TE_+}$ scores are slightly more dispersed than $\alpha_{\TA_+}$. This is reasonable because attack tactics are one level of abstraction higher than attack techniques. Hence, $\alpha_{\TE_+}$ exhibits greater sensitivity in measurement. 
Nevertheless, 55\% (60 out of the 108) attacks have both $\alpha_{\TA_+}$ and $\alpha_{\TE_+}$ scores of 0.5 or less. This leads to the following insight: 
\begin{insight}
Space cyber attacks of average sophistication can be successful (i.e., space cyber defenses have yet to eliminate the ``low hanging fruits'' that benefit attackers).
\end{insight}



\noindent{\bf How many attacks would have been stopped by using simple countermeasures?}
We observe that 48 (out of the 108) attacks employ techniques to establish an on-path attack position and that the $\alpha_{\TE_+}$ scores for these attacks are 0.4 or less. Although these attacks typically target the user segment, they leverage the weakness of the link segment. This leads to the insight:

\begin{insight}
Proper security of the link segment between the space and user segments (e.g., using cryptography) could have thwarted nearly half 
of the 108 space cyber attacks. 
\end{insight}

However, this insight does not imply the ease of employing cryptosystems for space infrastructures due to various barriers that need to be overcome (e.g., low compute capabilities onboard satellites, 
incorporation of legacy systems). Yet, it does point to the urgency to overcome such barriers.

We observe that 23 (out of the 108) attacks employ cyber social engineering techniques; among the 23 attacks, 20 have consequences in the ground segment where $\vec{g}_\G$ scores range from 0.4 to 0.7,
while their $\alpha_{\TE_+}$ scores
range from 0.4 to 0.9. The average sophistication score of all the social engineering
attacks 
is 0.3, meaning that mitigations against unsophisticated cyber social engineering attacks can thwart sophisticated space cyber attacks with significant consequences.
However, this observation is made at the attack technique level of abstraction, while noting that sophistication of cyber social engineering attack techniques is highly nuanced at the 
procedure level \cite{longtchi2024quantifying}. 
This leads to the following insight:

\begin{insight}
Traditional IT security controls against cyber social engineering attacks could have thwarted 32\% (20/63) of the cyber attacks that compromise the ground segment.
\end{insight}




\subsubsection{Attack Likelihood Analysis}
As mentioned above, the 108 space cyber attacks collectively employ 14 attack tactics and 107 attack techniques.
We compute the likelihood that a $\USCKC$ corresponds to the ground-truth of the corresponding incident, denoted by $L(\USCKC)\in (0,1]$, as follows. First, we apply our domain knowledge to attain measurements for the likelihood of each attack technique involved in the $\{\USCKC\}$ resulting from Algorithm \ref{alg:data extrapolation}.
For this purpose, and as depicted in Figure \ref{fig:sophistication-likelihood-per-at}, we assign each of the 107 attacks techniques a likelihood score from 0 to 1, while considering 0.2 the average likelihood 
because there are many obstacles attackers must overcome, such as fulfilling the attack technique's data and access requirements.
We apply the measurement to each $te \in {\TE}$ corresponding to a specific $\USCKC$ (i.e., one entry in the dataset), 
and use Definition \ref{definition:kill-chain-likelihood} to compute the likelihood 
each 
$\USCKC \in \{\USCKC\}$, namely $L(\USCKC)$. 
Then, we compute the overall likelihood of all 
$\USCKC$s for each attack,
leading to $L(\{\USCKC\})$.
For example, a 
$\USCKC$ for a Data Corruption/Interception attack against a NASA network in the ground segment in 2019 contains the following attack techniques (with accompanying likelihood scores): Valid Accounts (0.22), Exploitation of Remote Services (0.09), Indicator Removal (0.05), and Resource Hijacking (0.25). The likelihood of the $\USCKC$ 
is bounded by the least likely technique, leading to $L(\USCKC)=0.05$. Given that $L(\USCKC)=0.05$ is the highest likelihood score among all the probable and plausible $\USCKC$s of this attack, we have
$L(\{\USCKC\}) = 0.05$.

\begin{figure*}[!htbp]
\centering
\includegraphics[width=1.8\columnwidth]{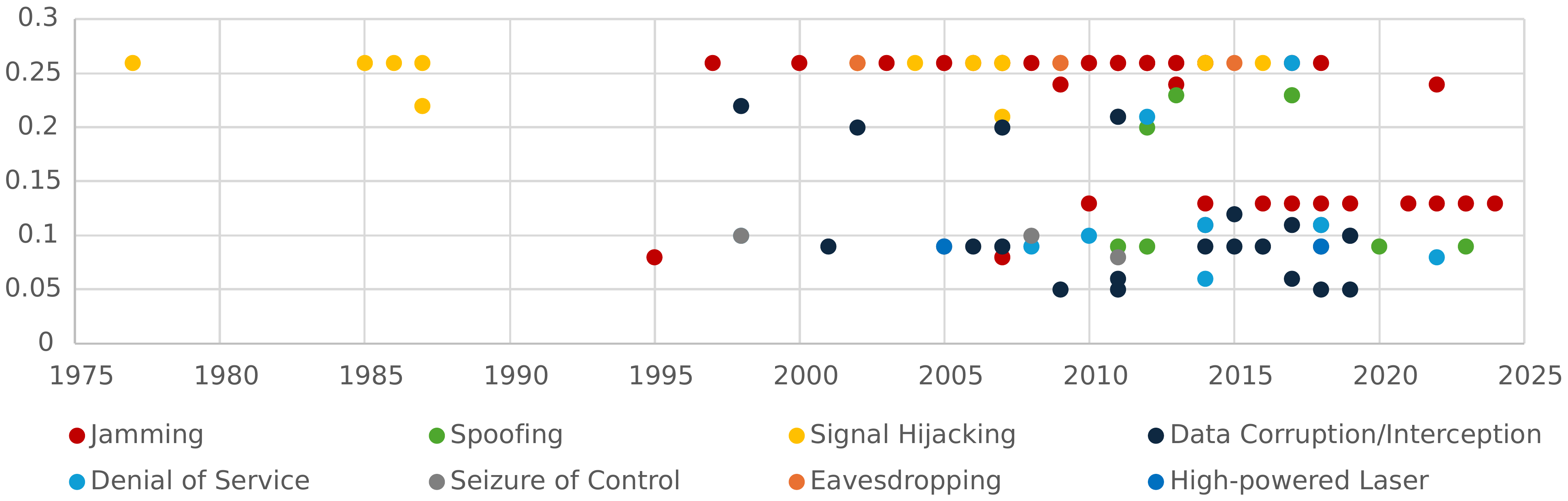}
\vspace{-1em}
\caption{$\USCKC$ likelihood of the 108 attacks over time.}
\label{fig:likelihood_over_time}
\end{figure*}

Figure~\ref{fig:likelihood_over_time} depicts $L(\{\USCKC\})$ of the 108 space cyber attacks. When considering $L(\{\USCKC\})$ in relation to attack sophistication $\alpha_{\TE_+}$, we observe that
attacks that have a higher $L(\{\USCKC\})$ often have a lower $\alpha_{\TE_+}$. For example, out of the 57 attacks where $L(\{\USCKC\}) \geq 0.2$, 51 of the 57 attacks have $\alpha_{\TE_+} \leq 0.5$. This leads to the following insight:

\begin{insight}
\label{insight:sophistication vs success}
Less sophisticated unified space cyber kill chains are more widely used
by real-world attackers.
\end{insight}

Insight \ref{insight:sophistication vs success} suggests attackers often seek ``low hanging fruits.'' For example, in the 2014 Data Corruption/Interception attack against JPL where malware was used to compromise a data processing server, 
the attacker uses the Privilege Escalation tactic, for which we extrapolate 3 ATT\&CK techniques: T1611 (Escape to Host), T1631 (Process Injection), and T1078 (Valid Accounts). 
Assuming the attacker has these three techniques, the most readily executable technique is T1078 (``low hanging fruit''), which has the highest $L(te)=.22$ among the 3, while also being among the least
sophisticated from the 107 $te$'s that are used by the 108 attacks. Our analysis is corroborated by \cite{LoganMilitaryCyberAffairs2024}, which shows that
Advanced Persistent Threats (APT) in real-world cyber attacks tend to employ less sophisticated attack techniques (e.g., APT28 employs T1078 all 6 times in the 6-step attack against the Democratic Congressional Campaign Committee).

\section{LIMITATIONS}\label{sec:limits}

The present study has several limitations. First, we address the missing-data problem in a  manual fashion, rather than an automated fashion. 
Since we leverage our domain expertise to manually extrapolate attack tactics, attack techniques, space cyber kill chains, and space cyber attack campaigns, the process is inevitably subjective even though we strive to be as objective as possible. 
Future research should investigate automated and objective methods for this process, potentially Large Language Models.

Second, our metrics can be refined. For example, the attack consequence  metrics, except those associated with Link Segments, are geared toward  {\em availability} because the raw dataset lacks enough information about what kinds of confidential data are processed in these space infrastructures; as a matter of fact, only 3
out of the 108 attack descriptions contain such information to some extent.
Future research should refine the metrics to accommodate, for example, data-specific {\em confidentiality} and {\em integrity}. For this purpose, one source of inspiration would be the CVSS scores of vulnerabilities \cite{CVSSv3} because their exploitation  causes attacks. However, one cannot simply adopt the CVSS scores until after figuring out what software vulnerabilities enabled space cyber attacks, as vulnerabilities are not documented in raw incident reports. 

Third, we assume the measurements of the ``building-block'' metrics are given as input. While reasonable because of the focus of the present study, it is important to investigate how to obtain these measurements, which may require a community effort. 
In our case study, we use our own subjective measurements.


\section{RELATED WORK}\label{sec:related_works}

We divide related prior studies into four categories: analyzing space infrastructure incidents in the real world, analyzing threats against space infrastructures in an abstract model, employment of IT cybersecurity frameworks to space infrastructures, and cybersecurity metrics.


There are studies on analyzing space-related incidents \cite{falco2021security,
pavur2022building,fritz2013satellite,boschetti2022space, pavur2020tale, soesanto2021terra}.
For example, \cite{pavur2022building} considers incidents in terms of the payload, signal, and ground aspects; \cite{fritz2013satellite} provides narrative descriptions concerning NASA, jamming, hijacking, and control attack categories; \cite{falco2021security} analyzes 1,847 space-related incidents according to their risk taxonomy for space. 
We leverage the taxonomy presented in \cite{falco2021security} to categorize the incidents compiled within this paper (e.g., hijacking, control, or DoS type of incidents).
There are also studies on individual real-world incidents \cite{boschetti2022space, pavur2020tale, soesanto2021terra}. 
By contrast, we present the most comprehensive real-world space cyber attacks with 108 attacks. These 108 attacks are complementary to the 25 attacks described in the academic literature \cite{XuS&P2025}. Note that only one class of real-world space cyber attacks, namely {\em data interception}, is not studied in academic literature, perhaps because {\em data interception} should have been prevented by security design.

A 2021 report \cite{harrison2021space} describes threats to space, especially physical and electronic capabilities of nation-states. 
However, it provides no specific cyber capabilities against space infrastructures, and neither do the other similar studies  \cite{thangavel2022understanding, falco2021security, soesanto2021terra, tedeschi2022satellite}. 
There are studies on threat models for space applications, such as leveraging LEO constellations to jam GEO targets \cite{rawlins2022death}, creating attack trees against CubeSats \cite{falco2021cubesat}, demonstration of command injection via a software-defined radio \cite{lin2022defending}, characterizing the transmission layer's susceptibility to Eavesdropping \cite{richardson2022ensuring}, leveraging the ATT\&CK framework \cite{ormrod2021cyber}, nanosatellites as attack platforms \cite{pavur2021same}, and space cyber risk management systems and properties \cite{XuIEEECSR2025-NRS,XuIEEECSR2025-Properties}.
Moreover, we refer to \cite{XuS&P2025} for a systematization of academic space cybersecurity research. By contrast, we analyze real-world cyber attacks via the notion of $\USCKC$ and metrics.

There are studies on the application of IT cybersecurity frameworks to the space domain, such as for defined space mission areas with cybersecurity overlaid upon them \cite{cunningham2016towards, zatti2017protection, vivero2013space}, a more hybrid approach combining both mission areas and explicit threat models \cite{book2006security}, and applying pre-existing IT controls to map to space infrastructures \cite{knez2016lessons, young2017commercial, vera2016cyber, rose2022building}. These studies help consider space cybersecurity holistically. However, they typically treat cybersecurity requirements with abstract and generalized concepts. Our present study applies threat-centric cybersecurity taxonomies to real-world space-related incidents.

Finally, there are very few studies on cyber attack sophistication metrics for space-related incidents. One study referenced an aerospace resiliency framework which provides categories where cybersecurity metrics could be developed \cite{thangavel2022understanding}. \cite{tedeschi2022satellite} provides performance metrics of physical layer defense. \cite{Pendleton16} provides a survey of current and proposed cybersecurity metrics that is most useful for our study, especially in its discussion of measuring attacks, evasion techniques, evasion capability, obfuscation sophistication, and power of targeted attacks. We leverage these studies to define new sophistication metrics and apply them to the space-related cybersecurity incidents. 
Moreover, our framework is innovative and includes metrics that have not appeared in the literature \cite{Pendleton16,XuSTRAM2018ACMCSUR,XuAgility2019,
XuSciSec2021SARR}

\section{CONCLUSION}\label{sec:conclusion}
We presented an initial study on characterizing cyber attacks against space infrastructures, which are extracted from 4 public datasets and other sources across the Internet documenting broader space-related incidents. 
We proposed an innovative framework with precisely defined metrics, while addressing the missing-data problem with conceptual algorithms because these real-world attacks are poorly documented. 
We prepared the first dataset of cyber attacks against space infrastructures that include hypothetical but plausible attack details. By applying the framework to the dataset, we drew a number of insights. The limitations represent outstanding open problems for future research.





\end{document}